\documentclass[reprint,amsmath,amssymb,aps,prb,superscriptaddress,longbibliography]{revtex4-1}
\pdfoutput=1

\usepackage{hyperref,url}
\usepackage{amsmath}
\usepackage{amsfonts}
\usepackage[capitalise]{cleveref}
\usepackage{placeins}
\hypersetup{breaklinks,colorlinks=true,linkcolor=blue,citecolor=blue,urlcolor=blue,filecolor=blue}
\usepackage{graphicx}
\usepackage{dcolumn}
\usepackage[version=3]{mhchem}
\usepackage[english]{babel}
\usepackage{filecontents}
\usepackage{makecell}
\usepackage{float}
\usepackage{graphicx}
\usepackage[outercaption]{sidecap}    
\usepackage{courier}
\usepackage{amssymb,amsmath,amsthm,mathrsfs,comment,afterpage,accents}
\usepackage{mathtools}
\usepackage{braket}
\usepackage{mathrsfs}
\usepackage{courier}
\usepackage{color}
\usepackage{subdepth}
\usepackage{tablefootnote}
\usepackage{multirow}
\usepackage{booktabs}
\usepackage{siunitx}
\usepackage[normalem]{ulem}

\newcolumntype{?}{!{\vrule width 1pt}}

\newcommand\mat\mathbf
\newcommand\tr{\operatorname{tr}}

\usepackage{physics}
\newcommand{\eq}[1]{Eq.~\hyperref[eq:#1]{(\ref*{eq:#1})}}
\newcommand{\citen}[1]{Ref.~\citenum{#1}}
\renewcommand{\sec}[1]{\hyperref[sec:#1]{Section~\ref*{sec:#1}}}
\newcommand{\app}[1]{\hyperref[app:#1]{Appendix~\ref*{app:#1}}}
\newcommand{\tab}[1]{\hyperref[tab:#1]{Table~\ref*{tab:#1}}}
\newcommand{\fig}[1]{\hyperref[fig:#1]{Figure~\ref*{fig:#1}}}

\newcommand{\stab}{\mathrm{stab}}
\newcommand{\CNOT}{\text{CNOT}}
\newcommand{\CX}{\text{CX}}

\newcommand{\CZ}{\text{CZ}}
\newcommand{\transpose}{\mathrm{T}}

\newcommand{\Google}{\affiliation{Google Quantum AI, Mountain View, CA, USA}}
\newcommand{\Columbia}{\affiliation{Department of Chemistry, Columbia University, New York, NY, USA}}
\newcommand{\Cal}{\affiliation{Berkeley Quantum Information \& Computation Center, University of California, Berkeley, CA, USA}}

\begin{document}

\author{William J. Huggins} 
\thanks{These two authors contributed equally; corresponding author: whuggins@google.com}
\Google
\author{Bryan A.~O'Gorman} 
\Cal
\author{Nicholas C.~Rubin} 
\Google
\author{David R.~Reichman}
\Columbia
\author{Ryan Babbush}
\Google
\author{Joonho Lee}
\thanks{These two authors contributed equally; corresponding author: linusjoonho@gmail.com}
\Columbia\Google
\title{Unbiasing Fermionic Quantum Monte Carlo with a Quantum Computer}


%
%
%
\begin{abstract}
Many-electron problems pose some of the greatest challenges in computational science, with important applications across many fields of modern science. Fermionic quantum Monte Carlo (QMC) methods are among the most powerful approaches to these problems. However, they can be severely biased when controlling the fermionic sign problem using constraints, as is necessary for scalability.
Here we propose an approach that combines constrained QMC with quantum computing tools to reduce such biases. We experimentally implement our scheme using up to 16 qubits in order to unbias constrained QMC calculations performed on chemical systems with as many as 120 orbitals. These experiments represent the largest chemistry simulations performed on quantum computers (more than doubling the size of prior electron correlation calculations), while obtaining accuracy competitive with state-of-the-art classical methods. Our results demonstrate a new paradigm of hybrid quantum-classical algorithm, surpassing the popular variational quantum eigensolver in terms of potential towards the first practical quantum advantage in ground state many-electron calculations.
\end{abstract}
\maketitle 

 {\it Introduction}. 
An accurate solution of the Schr{\"o}dinger equation for the ground state of many-electron systems is of critical importance across many fields of modern science.\cite{Friesner2005May,Helgaker2008Aug,Cao2019Oct,Bauer2020Nov} The complexity of this equation seemingly grows exponentially with the number of electrons in the system.  This fact has greatly hindered progress towards an efficient means of accurately calculating ground state quantum mechanical properties of complex systems. Over the last century, a substantial research effort has been devoted to the development of new algorithms for the solution of the many-electron problem.  Currently, all available general-purpose methods can be grouped into two categories: (1) methods which scale exponentially with system size while yielding numerically exact answers and (2) methods whose cost scales polynomially with system size but which are approximate by construction.  Approaches of the second category are currently the only methods that can feasibly be applied to large systems.  The accuracy of solutions obtained by these methods may be unsatisfactory and is nearly always difficult to assess.

Quantum computing has arisen as an alternative paradigm for the calculation of quantum properties that may complement and potentially surpass classical methods in terms of efficiency.\cite{Feynman1982,Lloyd1996}  While the ultimate ambition of this field is to construct a universal fault-tolerant quantum computer,\cite{Shor1996} the experimental devices of today are limited to Noisy Intermediate-Scale Quantum (NISQ) computers.\cite{Preskill2012}
NISQ algorithms for the computation of ground states have largely centered around the variational quantum eigensolver (VQE) framework,\cite{Peruzzo2013,McClean2015} which necessitates coping with optimization difficulties, measurement overhead, and circuit noise. As an alternative, algorithms based on imaginary time evolution have been put forward that, in principle, avoid the optimization problem.\cite{McArdle2019Sep,Motta2020Feb} However, due to the non-unitary nature of imaginary time evolution, one must resort to optimization heuristics in order to achieve reasonable scaling with system size. New computational strategies which avoid these limiting factors may help to enable the first practical quantum advantage in fermionic simulations.  In this work, we propose and experimentally demonstrate a new class of quantum-classical hybrid algorithms that offers a different route to addressing these challenges. We do not attempt to represent the ground state wavefunction using our quantum processor, choosing instead to use it to guide a quantum Monte Carlo calculation performed on a classical coprocessor.  Our experimental demonstration surpasses the scale of all prior experimental works on the quantum simulation of chemistry.\cite{kandala2017hardware,nam2020ground,google2020hartree}

\begin{figure*}[!ht]
\centering
\includegraphics[width=160mm]{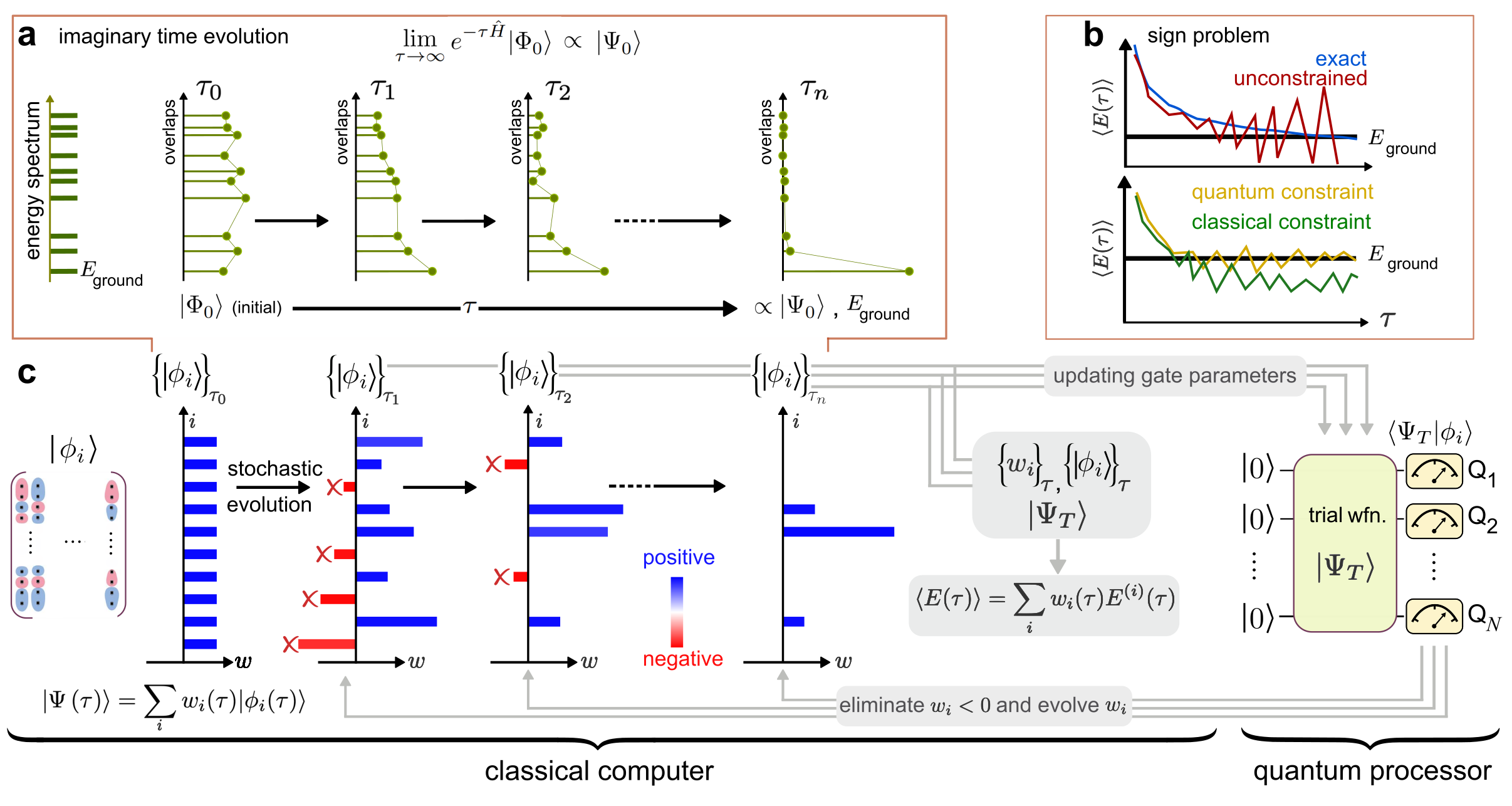}
\caption{(a) Depiction of the imaginary time evolution which shows an exponential convergence to the ground state as a function of imaginary time $\tau$. (b) Illustration of the fermionic sign problem. (b, top) Exact deterministic imaginary time evolution and an unconstrained QMC calculation which is exact on average but the signal-to-noise ratio in the average energy $\langle E(\tau)\rangle$ diverges in $\tau$ due to the sign problem. (b, bottom) Constrained QMC calculations with classical and quantum constraints. The use of quantum constraint can help to reduce the bias that is non-negligible when using the classical constraint. (c) Overview of the quantum-classical hybrid QMC (QC-QMC) algorithm. Stochastic wavefunction samples are represented as $\{|\phi_i\rangle\}_\tau$ (depicted as a matrix manageable to store classically) which are evolved in time along with associated weights $\{w_i\}_\tau$. Throughout the time evolution, queries to the quantum processor about the overlap value between the quantum trial wavefunction $|\Psi_T\rangle$ and a stochastic wavefunction sample $\{|\phi_i\rangle\}_\tau$ are made while updating the gate parameters to describe $\{|\phi_i\rangle\}_\tau$. Our quantum processor uses $N$ qubits to efficiently estimate the overlap which is then used to evolve $w_i$ and to discard stochastic wavefunction samples with $w_i< 0$. Finally, observables such as $\langle E(\tau)\rangle$ are computed on the classical computer by only making overlap queries to the quantum processor (see \cref{app:ovlploc}).
}
\label{fig:fig1}
\end{figure*}

{\it Theory and algorithms}. 
Quantum Monte Carlo (QMC) approaches\cite{Acioli1997May,foulkes_rmp} target the exact ground state $|\Psi_0\rangle$ of a many-body Hamiltonian, $\hat{H}$, via imaginary time evolution of an initial state $|\Phi_0\rangle$ with a non-zero overlap with $|\Psi_0\rangle$:
\begin{equation}
|\Psi_0\rangle \propto \lim_{\tau\rightarrow\infty} |\Psi(\tau)\rangle,\quad
|\Psi(\tau)\rangle \equiv e^{-\tau \hat{H}} | \Phi_0\rangle,
\label{eq:imag}
\end{equation}
where $\tau$ is imaginary time and $|\Psi(\tau)\rangle$ denotes the time-evolved wavefunction from $|\Phi_0\rangle$ by $\tau$ (see \cref{fig:fig1}(a)).
In QMC, the imaginary-time evolution in \cref{eq:imag} is implemented stochastically, which can enable a polynomial-scaling algorithm to sample an estimate for the exact ground state energy by avoiding the explicit storage of high dimensional objects such as $\hat{H}$ and $|\Psi_0\rangle$. The ground state energy, $E_\text{ground} =E(\tau=\infty)$, is estimated from averaging a time series of $\{\langle E(\tau) \rangle\}$, given by a weighted average over $M$ statistical samples,
\begin{equation}
\langle E(\tau) \rangle = \sum_{i=1}^M w_i(\tau) E^{(i)}(\tau),
\label{eq:energy0}
\end{equation}
where $E^{(i)}(\tau)$ is the $i$-th statistical sample for the energy and $w_i(\tau)$ is the corresponding normalized weight for that sample at imaginary time $\tau$.
While formally exact, such a stochastic imaginary time evolution algorithm will generically run into the notorious fermionic sign problem,\cite{Troyer2005May} which manifests due to alternating signs in the weights of each statistical sample used in \cref{eq:energy0}. In the worst case, the fermionic sign problem causes the estimator of the energy in \cref{eq:energy0} to have exponentially large variance (see \cref{fig:fig1}(b) top), necessitating that one averages exponentially many samples to obtain a fixed precision estimate of observables such as the ground state energy. Accordingly, exact, unbiased QMC approaches are only applicable to small systems\cite{Blankenbecler1981Oct,Chang2015Apr} or those lacking a sign-problem.\cite{Li2019Mar}

The sign problem can be controlled to give an estimator of the ground state energy with polynomially bounded variance by imposing constraints on the imaginary time evolution of each statistical sample represented by a wavefunction, $|\phi_i(\tau)\rangle$. These constraints (which include prominent examples such as the fixed node\cite{Moskowitz1982Jul,foulkes_rmp} and phaseless approximations\cite{zhang_cpmc,zhang_phaseless}) are imposed by the use of trial wavefunctions ($|\Psi_T\rangle\rangle$), and the accuracy of constrained QMC is wholly determined by the choice of the trial wavefunction (see \cref{fig:fig1}(b) bottom). 
Such constraints necessarily introduce a potentially significant bias in the final ground state energy estimate which can be removed in the limit that the trial wavefunction approaches the exact ground state.  

Classically, computationally tractable options for trial wavefunctions are limited to states such as a single mean-field determinant (e.g.~a Hartree-Fock state), a linear combination of mean-field states, a simple form of the electron-electron pair (two-body) correlator (usually called a Jastrow factor) applied to mean-field states, or some other physically motivated transformations applied to mean-field states such as backflow approaches.\cite{Becca2017Nov}
On the other hand, any wavefunction preparable with a quantum circuit is a candidate for a trial wavefunction on a quantum computer, including 
more general two-body correlators. 
These trial wavefunctions will be referred to as ``quantum'' trial wavefunctions. 

To be more concrete, there is currently no efficient classical algorithm to estimate (to additive error) the overlap between $|\phi_i(\tau)\rangle$ and certain complex quantum trial wavefunctions $|\Psi_T\rangle$ such as unitary coupled-cluster with singles and doubles\cite{Bartlett1989Feb} or the multiscale entanglement renormalization ansatz,\cite{Evenbly2015Nov} even when $|\phi_i(\tau)\rangle$ is  simply a computational basis state or a Slater determinant. Since quantum computers can efficiently approximate $\langle \Psi_T|\phi_i(\tau)\rangle$, there is a potential quantum advantage in this task as well as its particular use in QMC. This offers a different route towards quantum advantage in ground-state fermion simulations as compared to VQE, which instead seeks an advantage in the variational energy evaluation. We expand on this discussion of quantum advantage in \app{quantumadvantage}. We also note that VQE may be used to generate a sophisticated trial wavefunction which alone would not be sufficient to achieve high accuracy, but might offer quantitative accuracy and even quantum advantage when used as a trial wavefunction in our approach.

Our quantum-classical hybrid QMC algorithm (QC-QMC) utilizes quantum trial wavefunctions while performing the majority of imaginary time evolution on a classical computer, and is summarized in \cref{fig:fig1}(c).
In essence, on a classical computer one performs imaginary time evolution for
each wavefunction statistical sample, $|\phi_i(\tau)\rangle$, 
and collects observables such as the ground state energy estimate, $E^{(i)}(\tau)$.
During this procedure, a constraint associated with the quantum trial wavefunction is imposed to control the sign problem.
To perform the constrained time evolution, the only quantity that needs to be calculated on the quantum computer is the overlap between the trial wavefunction, $|\Psi_T\rangle$, and the statistical sample of the wavefunction at imaginary time $\tau$, $|\phi_i(\tau)\rangle$.

\begin{figure*}[!ht]
    \centering
\includegraphics{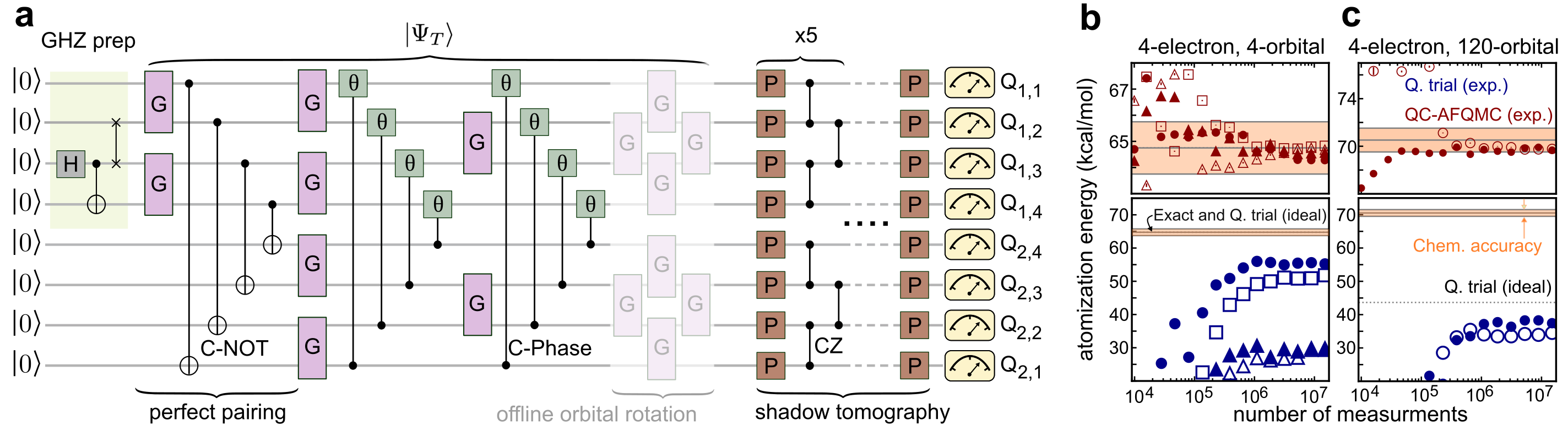}
    \caption{(a) Experimental circuit used for the 8-qubit ${\rm H}_4$ experiment over a 2x4 qubit grid (from $\text{Q}_{1,1}$ to $\text{Q}_{2,1}$) on the Sycamore quantum processor.\cite{Arute2019} In the circuit diagram, H denotes the Hadamard gate, G denotes a Givens rotation gate (generated by ${\rm XX} + {\rm YY}$), P denotes a Pauli gate, and $|\Psi_T\rangle$ denotes the quantum trial wavefunction. Note that the ``offline'' orbital rotation is not present in the actual quantum circuit because they can be efficiently handled via classical post-processing as discussed in \cref{app:trial}. (b) and (c): Convergence of the atomization energy of \ce{H4} as a function of the number of measurements. (b) a minimal basis set (STO-3G) with four orbitals total from four independent experiments with different sets of random measurements
    and (c) a quadruple-zeta basis set (cc-pVQZ) with 120 orbitals total from two independent experiments. The different symbols in (b) and (c) show independent experimental results. Note that the ideal (i.e., noiseless) atomization energy of quantum trial in (b) is exactly on top of the exact one. Further note that the quantum resource used in (c) is 8-qubit, but as shown in \cref{app:virtual}, our algorithm allows for adding ``virtual'' electron correlation in a much larger Hilbert space. Top panels of (b) and (c) magnifies the energy range near the exact answer.}
    \label{fig:fig2}
\end{figure*}

In this work, we estimate the overlap between the trial wavefunction and the statistical samples using a technique known as shadow tomography.\cite{Aaronson2020-ei,Huang2020-us}
Experimentally, this entails performing randomly chosen measurements of a reference state related to \(|\Psi_T\rangle\) prior to beginning the QMC calculation. 
In this formulation of QC-QMC, we emphasize that there is no need for the QMC calculation to iteratively query the quantum processor, despite the fact that the details of the statistical samples are not determined ahead of time.
By disentangling the interaction between the quantum and classical computer we avoid feedback latency, an appealing feature on NISQ platforms that comes at the cost of requiring potentially expensive classical post-processing (see \cref{app:vmc} for more details).
Furthermore, our algorithm naturally achieves some degree of noise robustness explained in \cref{app:noise_resilience} because the quantity that is directly is the ratio between overlap values, which is inherently resilient to the overlaps being rescaled by certain error channels.

While our approach applies generally to any form of constrained QMC, here we discuss an experimental demonstration of the algorithm that uses an implementation of QMC known as auxiliary-field QMC (AFQMC), which will be referred to as QC-AFQMC. In AFQMC, the phaseless constraint\cite{zhang_phaseless} is imposed to control the sign problem, and $|\phi_i(\tau)\rangle$ takes the form of a single Slater determinant in an arbitrary single-particle basis.
AFQMC has been shown to be accurate in a number of cases even with classically available trial wavefunctions;\cite{Zheng2017Dec,williams2020direct} however, the bias incurred from the phaseless constraint cannot be overlooked, as we discuss in detail below. Since a single determinant mean-field wavefunction is the most widely used classical form of the trial function for AFQMC, here we will use ``AFQMC'' to denote the use of AFQMC with mean-field trial wavefunction.

\begin{table}[h]
\begin{tabular}{c|c|c|c|c|c}
& Exact & AFQMC & CCSD(T) & Q. trial & QC-AFQMC \\ \hline
4-orbital &64.7 &62.9 & 59.6 & 55.2 &  64.3\\ \hline
120-orbital &70.5 &68.6 & 71.9 & 37.4 & 69.7
\end{tabular}
\caption {Atomization energy (kcal/mol) of \ce{H4} for quantum trial (Q. trial; experiment), AFQMC (classical), QC-AFQMC (experiment), CCSD(T) (classical ``gold standard''), and exact results for minimal (STO-3G; 4-orbital) and quadruple-zeta (cc-pVQZ; 120-orbital) bases. Both of these experiments use 8 qubits. The statistical error of AFQMC and QC-AFQMC is less than 0.05 kcal/mol and therefore not shown here.
Note that for QT and QC-AFQMC we picked an experiment done with a specific set of random measurements that are converged at 1.5$\times10^7$ measurements. As shown in \cref{app:comp}, QT results vary significantly run-to-run while QC-AFQMC results are nearly identical run-to-run (which showcases the noise resilience of QC-AFQMC).
}
\label{tab:h4}
\end{table}
{\it Results and discussion}.
The experiments in this work were carried out on Google's 54-qubit quantum processor, known as Sycamore.\cite{Arute2019}  The circuits were compiled using hardware-native CZ gates with typical error rates of $\approx 0.5\%$. \cite{Chen2021Feb}
As the first example, in \cref{fig:fig2}, we illustrate the quantum primitive used to perform shadow tomography on the \ce{H4} molecule in an 8-qubit experiment.
Our eight spin-orbital quantum trial wavefunction consists of a valence bond wavefunction known as a perfect pairing state\cite{Goddard1973Nov,Cullen1996-jr} and a hardware-efficient quantum circuit\cite{kandala2017hardware} with an offline single-particle rotation applied to this, which would be classically difficult to use as a trial wavefunction for AFQMC. The state preparation circuit in \cref{fig:fig2}(a) shows how this trial wavefunction can be efficiently prepared on a quantum computer. Similar state preparation circuits are used for the other chemical examples in this work.

\begin{figure*}[ht] 
\begin{minipage}[c]{0.58\textwidth}
    \includegraphics[width=\textwidth]{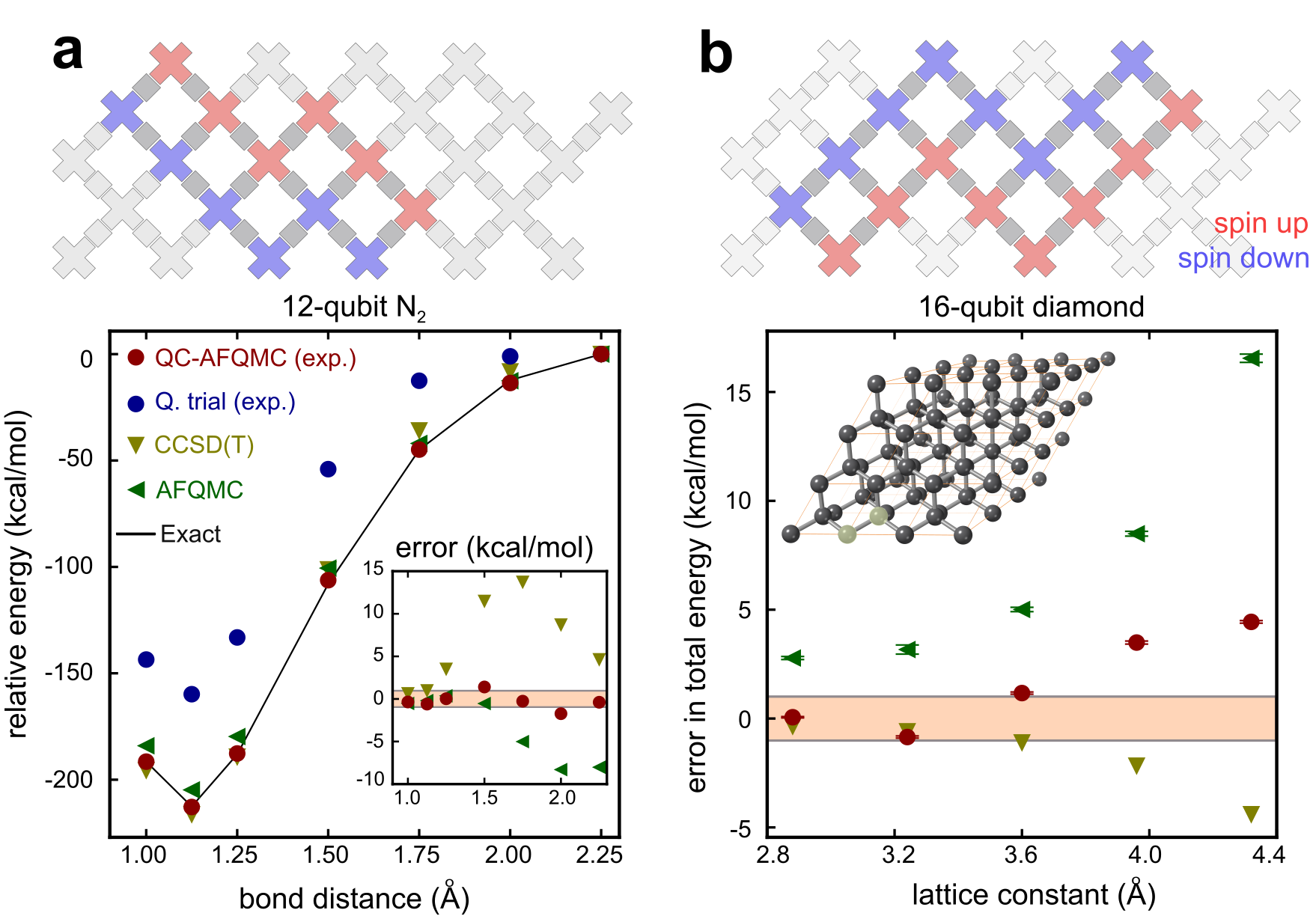}
  \end{minipage}\hfill
  \begin{minipage}[c]{0.35\textwidth}
\caption{(a, top) Circuit layout showing spin-up and spin-down qubits for the 12-qubit experiment. (a, bottom) Potential energy surface of \ce{N2} in a triple zeta basis set (cc-pVTZ\cite{Dunning1989}; 60-orbital). For clarity, the relative energies are shifted to zero at $2.25$\AA. Inset shows the error in total energy relative to the exact results in kcal/mol. The dash dotted line in the inset provides bounds for chemical accuracy (1 kcal/mol). Note that the variational energy of the quantum trial used here is outside the plotted energy scale. The statistical error bars of AFQMC and QC-AFQMC are not visible on this scale.
(b, top) Circuit layout showing spin-up and spin-down qubits for the 16-qubit experiment. 
(b, bottom) Error in total energy as a function of lattice constant of diamond in a double zeta basis (DZVP-GTH; 26 orbitals). The dash dotted line shows the bounds for chemical accuracy. Our quantum trial results are not visible on this energy scale. For high values of the lattice constant none of these methods achieve chemical accuracy but the use of the quantum trial still improves the AFQMC result. Inset shows a supercell structure of diamond where two highlighted atoms form the minimal unit cell. 
} 
\label{fig:fig3}
\end{minipage}
\end{figure*}

In this 8-qubit experiment, we consider \ce{H4} in a square geometry with side lengths of 1.23 \AA\: and its dissociation into four hydrogen atoms.  
This system is often used as a testbed for electron correlation methods in quantum 
chemistry.\cite{Paldus1993,Lee2019Jan} We perform our calculations using two Gaussian basis sets: the minimal (STO-3G) basis\cite{hehre1969self} and the correlation consistent quadruple-zeta (cc-pVQZ) basis.\cite{Dunning1989} The latter basis set is of a size and accuracy required to make a direct comparison with laboratory experiments.  When describing the ground state of this system, there are two equally important, degenerate mean-field states. 
This makes AFQMC with a single mean-field trial wavefunction highly unreliable. In addition, a method often referred to as a ``gold standard'' classical approach (coupled-cluster with singles, doubles, and perturbative triples, CCSD(T)\cite{Raghavachari1989May}) also performs poorly for this system.

In \cref{tab:h4}, the difficulties of AFQMC and CCSD(T) are well illustrated by comparing their atomization energies with exact values in two different basis sets. Both approaches show errors that are significantly larger than ``chemical accuracy'' (1 kcal/mol). 
The variational energy of the quantum trial reconstructed from experiment has a bias that can be as large as 33 kcal/mol. The noise on our quantum device makes the quality of our quantum trial far from that of the ideal (i.e., noiseless) ansatz as shown in \cref{fig:fig2}(b) and (c), resulting in an error as large as 10 kcal/mol in the atomization energy. Nonetheless, QC-AFQMC reduces this error significantly, and achieves chemical accuracy in both bases.

As shown in \cref{app:virtual}, for the larger basis set, we obtain a residual ``virtual'' correlation energy by using the quantum resources on a smaller number of orbitals to unbias an AFQMC calculation on a larger number of orbitals, with no additional overhead to the quantum computer. This capability makes our implementation competitive with state-of-the-art classical approaches. Similar virtual correlation energy strategies have been previously discussed within the framework of VQE,\cite{Takeshita2019} but unlike our approach, those strategies come with a significant measurement overhead.
To unravel the QC-AFQMC results on \ce{H4} further, we illustrate in \cref{fig:fig2}(b) and (c) the evolution of trial and QC-AFQMC energies as a function of the number of measurements made on the device. Despite the presence of significant noise within approximately $10^5$ measurements, QC-AFQMC achieves chemical accuracy while coping with a sizeable residual bias in the underlying quantum trial. 

Next, we move to a larger example, \ce{N2}, which requires a total of 12 qubits in our quantum experiment.
Here, a simpler quantum trial is used for QC-AFQMC by taking just the valence bond part of the wavefunction depicted in \cref{fig:fig2}(a).  We examine the potential energy surface of \ce{N2} from compressed to elongated geometries, which is another common benchmark problem for classical quantum chemistry methods.\cite{Siegbahn1983,Lee2019Jan}
In \cref{fig:fig3} (a), the QC-AFQMC result is shown for the calculations performed in a triple zeta basis (cc-pVTZ),\cite{Dunning1989} which corresponds to a 60-orbital or 120-qubit Hilbert space.  All examined methods, CCSD(T), AFQMC, and QC-AFQMC perform quite well near the equilibrium geometry, but CCSD(T) and AFQMC deviate from the exact results significantly as one stretches the bond distance. As a result, the error of  ``gold-standard'' CCSD(T) can be as large as 14 kcal/mol and the error of AFQMC with a classical trial wavefunction can be as large as -8 kcal/mol. The error in the QC-AFQMC computation ranges from -2 kcal/mol to 1 kcal/mol depending on the bond distance. Thus, while we do not achieve chemical accuracy with QC-AFQMC, we note that even with a very simple quantum trial wavefunction, we produce energies that are competitive with state-of-the-art classical approaches.

Lastly, we present a 16-qubit experiment result on the ground state simulation of a minimal unit cell (2-atom) model of periodic solid diamond in a double-zeta basis set (DZVP-GTH\cite{VandeVondele2007Sep}; 26 orbitals). While at this level of theory the model exhibits significant finite-size effects and does not predict the correct experimental lattice constant, we aim to illustrate the utility of our algorithm in materials science applications.  We emphasize that this is the largest quantum simulation of chemistry on a quantum processor to date. Previously, the largest correlated quantum simulations of chemistry involved half a dozen qubits or less\cite{kandala2017hardware} with more than an order of magnitude fewer two-qubit gates than is used here, while the largest mean-field calculation performed on a quantum computer involved a dozen qubits with fewer than half as many two-qubit gates.\cite{google2020hartree} We again use the simple perfect pairing state as our quantum trial wavefunction and  demonstrate the improvement over a range of lattice parameters compared with classical AFQMC and CCSD(T) in \cref{fig:fig3} (b). There is a substantial improvement in the error going from AFQMC to QC-AFQMC showing the increased accuracy due to better trial wavefunctions. Our accuracy is limited by the simple form of our quantum trial and yet we achieve accuracy nearly on par with the classical gold standard method, CCSD(T).

{\it Conclusion}. In summary, we 
proposed a scalable,
noise-resilient
quantum-classical hybrid algorithm
that seamlessly embeds a special-purpose quantum primitive
into an accurate quantum computational many-body method, namely QMC.
Our work offers an alternative computational strategy that effectively
unbiases fermionic QMC approaches
by leveraging state-of-the-art quantum information tools.
We have realized this algorithm for a specific QMC algorithm known as AFQMC, and
experimentally demonstrated its performance in experiments as large as 16-qubit on a NISQ processor, producing electronic energies that are competitive with state-of-the-art classical quantum chemistry methods.  Our algorithm also allows for incorporating the electron correlation energy outside the space that is handled by the quantum computer without increasing quantum resources or measurement overheads.
In \cref{app:quantumadvantage}, we discuss issues related to asymptotic scaling and the potential for quantum advantage in our algorithm, including the challenge of measuring wavefunction overlaps precisely. 
 While we have yet to achieve practical quantum advantage over available classical algorithms, the flexibility and scalability of our proposed approach in the construction of quantum trial functions, and its inherent noise resilience, promise a new path forward for the simulation of chemistry in the NISQ era and beyond. 


{\it Acknowledgements}. The authors thank members of the Google Quantum AI theory team and Fionn Malone for helpful discussions. BO is supported by a NASA Space Technology Research Fellowship.  The quantum hardware used for this experiment was developed by the Google Quantum AI hardware team, under the direction of Anthony Megrant, Julian Kelly and Yu Chen. Theoretical foundations for device calibrations were provided by the physics team lead by Vadim Smelyanskiy. Initial data collection was enabled by cloud access to these devices as part of Google Quantum AI's Quantum Computing Service Early Access Program. Pedram Roushan and Charles Neill from the Google team helped to execute the experiment on hardware and design figures.

{\it Note}. The code and data for this study are available from the corresponding authors upon request and will be made available publicly in the future. After this work was nearly complete, a theory paper by Yang et al. appeared on arXiv,\cite{Yang2021Jun} describing a quantum algorithm for assisting real time dynamics with unconstrained QMC.

{\it Author contributions}.
JL conceived of the quantum-classical hybrid QMC algorithm, performed QMC calculations, and with contribution from others drafted the manuscript.
WJH proposed the use of shadow tomography and designed the experiment with contributions from others.
BO helped with theoretical analysis and the compilation of circuits.
NCR helped with the presentation of figures.
JL and RB managed the scientific collaboration. All authors participated in discussions, the writing of the manuscript, and the analysis of the data.

\bibliography{refs,ryan_mendeley}

\onecolumngrid

\appendix

\section{Technical Introduction}
Despite the tremendous advances made in theoretical chemistry and physics over the past several decades, problems with substantial electron correlation, namely effects beyond those treatable at the Hartree-Fock level of theory, still present great challenges to the field.\cite{Friesner2005May,Helgaker2008Aug,Cao2019Oct,Bauer2020Nov}
Electron correlation effects play a central role in many important situations, ranging from the treatment of transition-metal-containing systems to the description of chemical bond breaking.  Reaching so-called ``chemical accuracy'' (accuracy to within 1 kcal/mol) in such applications  is the holy grail of quantum chemistry, and is a goal which no single method can currently reliably and scalably achieve. 

Among electronic structure methods, projector quantum Monte Carlo (QMC) has proven to be among the most accurate and scalable. QMC implements imaginary-time evolution of a quantum state with stochastic sampling
and can produce unbiased ground state energies when the fermionic sign problem is absent, for example in cases with particle-hole symmetry.
Widely used QMC methods include diffusion Monte Carlo (DMC), Greens function Monte Carlo (GFMC), and auxiliary-field QMC (AFQMC) approaches.\cite{Becca2017Nov}
Generally, chemical systems exhibit a fermionic sign problem and this significantly limits the applicability of QMC to small systems due to exponentially decreasing signal-to-noise ratio.\cite{Troyer2005May}  
Efficient QMC simulations for sizable systems are possible only with a constraint implemented in conjunction with a trial wavefunction on the imaginary-time trajectories, which at the same time introduces a bias in the final ground state energy estimate.

The accuracy of QMC simulations is, therefore, wholly determined by the quality of the trial wavefunction.
In cases where strong electron correlation is not present, using a simple single Slater determinant trial wavefunction
obtained from a mean-field (MF) approach leads to accurate approximate ground state energies from QMC.
However, for cases where MF wavefunctions are qualitatively wrong, 
one must resort to other alternatives. 
The form of wavefunction must be simple enough to evaluate the projection onto a working QMC basis in an efficient manner. 
The QMC basis takes the form of real-space points in DMC, occupation vectors in GFMC, and non-orthogonal Slater determinants in AFQMC.
The projection onto the QMC basis often scales exponentially with system size for coupled-cluster states and tensor-product states such as matrix product states. Trial wavefunctions consisting of a linear combination of determinants have been widely used due to the simple evaluation of the projection in this case. However, obtaining an accurate linear combination of determinants scales poorly because the number of important determinants generically scales exponentially with system size.
Given these facts, there is a need for a new paradigm that allows for more flexible choices of trial wavefunctions which can lead to more accurate QMC algorithms without losing their scalability.

In this work, we have proposed harnessing the power of quantum computers in performing a hybrid quantum-classical QMC simulation, which we refer to as the QC-QMC algorithm. The key observation that we exploit is that it is possible to perform the QMC basis projection for a wide range of wavefunctions in a potentially more efficient manner on quantum computers
than on classical computers.
This suggests that one may isolate the specific task of the projection from
the QMC algorithm and use quantum computers to perform this task and separately  communicate this information to a classical computer to continue the QMC calculation.
In principle the required quantity is straightforward to approximate using the Hadamard
test.\cite{Yu_Kitaev1995-zv}
However, because the QMC basis projection needs to be performed thousands of times for
a single QMC calculation, for Noisy Intermediate-Scale Quantum (NISQ) devices we propose using shadow tomography to characterize the trial wavefunction and
evaluate the projection such that the on-line interaction between the quantum and classical device no longer exists. 
This enables the exploration of the utility of quantum trial wavefunctions without concern for the challenges of tightly coupling high performance
classical computing resources with a NISQ device.
We demonstrate the usefulness and noise resilience of this approach by producing
accurate experiments through Google's Sycamore processor on prototypical
strongly correlated chemical systems such as \ce{H4} in a minimal basis and a
quadruple-zeta basis, as well as bond-breaking of \ce{N2} in a triple-zeta basis. We also studied a minimal unit cell model of diamond within a double zeta basis.

\section{Review of Projector Quantum Monte Carlo}
QMC methods are among the most accurate approximate electronic structure approaches, and they can be systematically improved with the use of increasingly sophisticated trial functions. 
Here, we summarize the essence of the algorithm and
discuss a specific QMC method which works in second-quantized space, namely auxiliary-field quantum Monte Carlo (AFQMC).
While we focus on developing a strategy tailored for AFQMC in this work,
the general discussion is not limited to AFQMC and should be applicable to QMC in general.
\subsection{Projector quantum Monte Carlo}
The essence of any projector QMC methods is that one computes
the ground state energy and properties 
via an imaginary-time propagation
\begin{equation}
|\Psi_0\rangle 
\propto
\lim_{\tau\rightarrow \infty}    
\exp \left(-\tau \hat{H}\right) |\Phi_0\rangle
= 
\lim_{\tau\rightarrow \infty}    
|\Psi(\tau)\rangle,
\label{eq:exact}
\end{equation}
where $\tau$ is the imaginary time,
$|\Psi_0\rangle$ is the exact ground state 
and $|\Phi_0\rangle$ is an initial starting wavefunction satisfying $\langle\Phi_0|\Psi_0\rangle \ne 0$. 
Without any further modification, this is an exact approach to the computation of the ground state wavefunction.
In practice, a deterministic implementation of \cref{eq:exact} scales exponentially with system size 
and therefore
one resorts to 
a stochastic realization of \cref{eq:exact}
for scalable simulations.
Such a stochastic realization
is typically referred to as
projector QMC.

Unfortunately, a direct implementation of \cref{eq:exact} via QMC 
suffers from
the infamous fermionic sign problem.\cite{Troyer2005May}
In first quantized QMC methods such as DMC, fermionic antisymmetry is not imposed explicitly. 
Such approaches require the imposition of the fermionic nodal structure using trial wavefunctions to compute the fermionic ground state. The use of an approximate nodal structure introduces a bias.
In second quantized QMC methods the sign problem manifests in a different way.
The statistical estimates from a second quantizated QMC method
exhibit variances that grow
exponentially with system size.
Therefore for simulations of large systems
no meaningful statistical estimates can be obtained.
It is then necessary to impose a constraint in the imaginary-time propagation to deal with the the sign problem and to obtain statistical efficiency.  An example of such a constraint is the ``phaseless" constraint in AFQMC (see below).
While such constraints introduce biases in the final estimates, 
rendering QMC approaches inherently approximate in practice, different constrained approaches will have relative strengths and weaknesses with respect to accuracy and flexibility.

\subsection{Auxiliary-field quantum Monte Carlo}
Auxiliary-field quantum Monte Carlo (AFQMC) 
is a projector QMC method that works in second-quantized space.\cite{motta_review}
Therefore, the sign problem in AFQMC
manifests
in growing variance in statistical estimates.
To impose a constraint in the imaginary-time propagation, 
it is natural to introduce a trial wavefunction 
that can be used in the importance sampling as well as the constraint.
This results in a wavefunction at imaginary time $\tau$ expressed as
\begin{equation}
|\Psi(\tau)\rangle
=
\sum_i
w_i (\tau)
\frac{| \phi_i (\tau)\rangle}{\langle \Psi_T | \phi_i (\tau)\rangle}
\label{eq:wfn}
\end{equation}
where $| \phi_i (\tau)\rangle$ is the wavefunction of the $i$-th walker, $w_i (\tau)$ is the weight of the $i$-th walker, and $|\Psi_T\rangle$ is some {\it a priori} chosen trial wavefunction.
From \cref{eq:wfn}, it is evident that the importance sampling is imposed based on the overlap between the walker wavefunction and the trial wavefunction.

Walker wavefunctions in \cref{eq:wfn} are almost always chosen to be  single Slater determinants
and the action of the imaginary propagation, $\exp(-\Delta \tau \hat{H})$, for a small time step $\Delta \tau$ in \cref{eq:exact}
transforms the walkers in such a way that they stay within the single Slater determinant manifold via the Hubbard-Stratonovich transformation.
This property is essential if the  computational cost is to grow only polynomially with system size, and is at the core of the AFQMC algorithm as well as that of another commonly used unconstrained (and therefore unbiased) projector QMC approach called the determinant QMC method.\cite{Blankenbecler1981Oct} 

While repeatedly applying the imaginary time propagator to the wavefunction, the AFQMC algorithm 
prescribes a particular way to update
the walker weight $w_i(\tau)$ in \cref{eq:wfn}.
In essence, it is necessary that all weights stay real and positive
so that the final energy estimator,
\begin{equation}
E(\tau)
=
\frac{\langle \Psi_T | \hat{H} | \Psi(\tau)\rangle}
{\langle \Psi_T | \Psi(\tau)\rangle}
=
\frac{\sum_i \omega_i E^{(i)}(\tau)}
{\sum_i \omega_i},
\label{eq:energy}
\end{equation}
has a small variance. Here,
$E^{(i)}(\tau)$ is so-called the local energy, which is defined as
\begin{equation}
E^{(i)}(\tau)
=
\frac{
\langle \Psi_T |\hat{H} |\psi_i (\tau)\rangle
}{\langle \Psi_T | \psi_i (\tau)\rangle}.
\label{eq:eloc}
\end{equation}
We note that \cref{eq:energy} is not a variational energy expression and is
commonly referred to as the ``mixed'' energy estimator in QMC.
The essence of the constraint is that one updates the $i$-th walker weight from $\tau$ to $\tau + \Delta\tau$ using
\begin{equation}
|S_i(\tau)| \times \text{max}(0, \cos\theta_i(\tau))
\end{equation}
where
\begin{equation}
S_i(\tau) = \frac{\langle
\Psi_T | \phi_i(\tau+\Delta\tau)
\rangle}{
\langle
\Psi_T | \phi_i(\tau)
\rangle},
\label{eq:ovl}
\end{equation}
and $\theta_i(\tau)$ is the argument of $S_i(\tau)$.
This is in a stark contrast with a typical importance sampling strategy which updates the walker weights using $S_i(\tau)$, which does not guarantee the positivity and reality of the walker weights.
If $|\Psi_T\rangle$ is exact, this constraint does not introduce any bias, but simply imposes a specific boundary condition on the imaginary propagation which can be viewed as a ``gauge-fixing'' of the wavefunction.
In practice, one does not have access to the exact $|\Psi_T\rangle$ and therefore can only compute an approximate energy whose accuracy wholly depends on the choice of $|\Psi_T\rangle$.
Such a constraint is usually referred to as the ``phaseless approximation'' in the AFQMC literature.


Currently, classically tractable trial wavefunctions that are commonly used are
either single determinant trials or take the form of a linear combination of determinants.\cite{Lee2021Feb,purwanto_cr2}
The former is very scalable (up to 500 electrons or so) but can be often inaccurate, especially for strongly correlated systems, while the latter is limited to a small number of electrons (16 or so) but can produce results that are very accurate even for strongly correlated systems.
The choice of the trial wavefunction renders AFQMC limited by the evaluation of \cref{eq:energy} and \cref{eq:ovl}.
If the computation of either one of these quantities scales exponentially with system size, the resulting AFQMC calculation will be exponentially expensive.

\section{Quantum-Classical Hybrid Auxiliary-Field QMC (QC-AFQMC) Algorithms}
In the main text, we presented the general philosophy of the QC-QMC algorithm and here we wish to provide more QC-AFQMC-specific details tailored to the experiments presented in this work.

From the perspective of QMC simulations, the main benefit of using a quantum computer
is to expand the range of available trial wavefunctions beyond what is efficient classically.
Namely, we seek a class of trial wavefunctions that are inherently more accurate than a single determinant trial while bypassing the difficulty of variational optimization on the quantum computer.  Among the set of possible trial functions, we are interested in using wavefunctions for which no known polynomial-scaling classical algorithm exists for the exact evaluation of \cref{eq:energy} and \cref{eq:ovl}.
The core idea in the QC-AFQMC algorithm is that one can approximately measure \cref{eq:energy} and \cref{eq:ovl} on the quantum computer and implement the majority of the imaginary-time evolution classically. Our goal is provide a roadmap for quantum computers to apply polynomial-scaling algorithms for the evaluation of \cref{eq:energy} and \cref{eq:ovl} up to additive errors and thus ultimately to observe quantum advantage in some systems. 
This clearly separates subroutines into those that need to be run on quantum computers and those on classical computers.

\subsection{Quantum trial wavefunctions}\label{app:trial}

The specific trial functions of interest in this work are simple variants of so-called coupled-cluster (CC) wavefunctions.
In quantum chemistry, CC wavefunctions are among the most accurate many-body wavefunctions.\cite{bartlett_rmp} 
They are defined by an exponential parametrization,
\begin{equation}
|\Psi\rangle = e^{\hat{T}} | \psi_0\rangle,
\end{equation}
where $| \psi_0\rangle$ is a single determinant reference wavefunction and the cluster operator $\hat{T}$ is defined as
\begin{equation}
\hat{T}
=
\sum_{ai}t_{i}^aa_a^\dagger a_i
+
\sum_{ijab}t_{ij}^{ab}a_b^\dagger a_a^\dagger a_j a_i +\cdots.
\end{equation}
We use $\{i,j,k,\cdots\}$ to denote occupied orbitals and $\{a,b,c,\cdots\}$ for unoccupied orbitals.
$\hat{T}$ can be extended to include single excitations (S), double excitations (D), triple excitations (T) and so on. 
The resulting CC wavefunction is then systematically improvable by including higher-order excitations.
The most widely used version involves up to doubles and is referred to as CC with singles and doubles (CCSD).
There is no efficient algorithm for variationally determining the CC amplitudes, $\mathbf t$; however, there is an efficient projective way to determine these amplitudes and the energy, although the resulting energy determined by this procedure is not variational.
Such non-variationality manifests as a breakdown of conventional CC, although it has been suggested that
the underlying wavefunction is still qualitatively correct and the projective energy evaluation is partially responsible for this issue.\cite{van2000benchmark} 

Employing CCSD (or higher-order CC wavefunctions) within the AFQMC framework is difficult because the overlap between a CCSD wavefunction and an arbitary Slater determinant cannot be calculated efficiently without approximations. This is true for nearly all non-trivial variants of coupled cluster. Notably, there is currently no known efficient classical algorithm for precisely calculating wavefunction overlaps even for the cases of coupled cluster wevefunctions with a
limited set of amplitudes, such as generalized valence bond perfect-pairing (PP).\cite{Goddard1973Nov,Cullen1996-jr}
In QC-AFQMC, we can efficiently approximate the required overlaps of such wavefunctions by using a quantum computer to prepare a unitary version of CC wavefunctions or
approximations to them. By using CC wavefunctions that we
can obtain circuit parameters classically, we are able to avoid a costly variational optimization procedure on the quantum device.

The simplified CC wavefunction ansatz that we utilize in this work is the generalized valence bond PP ansatz.
This ansatz is defined as
\begin{equation}
\ket{\Psi_\text{PP}} = e^{\hat{T}_\text{PP}}e^{\hat{\kappa}}\ket{\psi_0},
\end{equation}
where the orbital rotation operator is defined as
\begin{equation}
\hat{\kappa}
=
\sum_{pq}^{N_\text{orbitals}}
(\kappa_{pq}^\uparrow - \kappa_{qp}^\uparrow) \hat{a}_{p_\uparrow}^\dagger \hat{a}_{q_\uparrow}
+
(\kappa_{pq}^\downarrow - \kappa_{qp}^\downarrow) \hat{a}_{p_\downarrow}^\dagger \hat{a}_{q_\downarrow},
\end{equation}
and the PP cluster operator is
\begin{equation}
\hat{T}_\text{PP} = \sum_i^{N_\text{pairs}}t_{i} \hat{a}^\dagger_{i^*_\uparrow}\hat{a}_{i_\uparrow}\hat{a}^\dagger_{i^*_\downarrow}\hat{a}_{i_\downarrow}.
\end{equation}
In this equation, each \(i\) is an occupied orbital and each \(i^*\) is the
corresponding virtual orbital that is paired with the occupied orbital $i$. We map the spin-orbitals of this wavefunction to qubits using the Jordan-Wigner transformation.
We note that the pair basis in $t_i$ is defined in the rotated orbital basis defined by the orbital rotation operator.

Due to its natural connection with valence bond theory which often provides a more intuitive chemical picture than does molecular orbital theory, the PP wavefunction has played an important role in understanding chemical processes.\cite{Goddard1973Nov}
Despite its exponential scaling when implemented exactly on a classical computer, PP in conjunction with AFQMC has been discussed previously; see~\citen{purwanto2008eliminating}. We will explore the scaling of the PP-based approach in classical AFQMC and QC-AFQMC in more in detail below because this wavefunction is used in all of our experimental examples (see \cref{app:quantumadvantage}). 

The PP wavefunction is known to provide insufficient accuracy for the ground state energy in many important examples.
This is
best illustrated
in systems where inter-pair correlation becomes important,
such as multiple bond breaking processes.\cite{Small2011Oct}
While there exist ways to incorporate inter-pair correlation classically,\cite{VanVoorhis2000Feb,Small2014May,Lee2018Dec}
in this work we focus on 
adding multiple layers of hardware-efficient operators to the PP ansatz.
There are two kinds of these additional layers that we have explored:
\begin{enumerate}
\item The first class of layers includes only density-density product terms of the form
\begin{equation}
e^{J_{ij}\hat{n}_i\hat{n}_j}.
\end{equation}
\item The second class includes only ``nearest-neighbor'' hopping terms between same spin ($\sigma$) pairs
\begin{equation}
e^{Q_{ij}\hat{a}_{i_\sigma}^\dagger \hat{a}_{j_\sigma} - Q_{ij}^*\hat{a}_{j_\sigma}^\dagger \hat{a}_{i_\sigma}}.
\end{equation}
\end{enumerate}
In both cases, the $i$ and $j$ orbitals are physically neighboring in the hardware layout.
We alternate multiple layers of each kind and apply these layers to the PP ansatz to improve the overall accuracy.
The efficacy of these layers varies with their ordering with the choice of the $i$,$j$ pairs. 
Lastly, we also employ a full single particle rotation at the end of the hardware-efficient layers. This last orbital rotation can be applied to 1-body and 2-body Hamiltonian matrix elements classically, so we do not have to implement this part on the quantum computer.
We refer this orbital rotation as ``offline orbital rotation'' as noted in \cref{fig:fig2}.
\ce{H4} was the only example where we went beyond the PP wavefunction.
When this type of hardware-efficient layers is used, we no longer have an efficient classical algorithm to optimize the wavefunction parameters.
In such cases, one can resort to the variational quantum eigensolver to obtain these parameters.
Nevertheless, in the case of \ce{H4}, the Hilbert space is small enough (4-orbital) that we still could optimize everything classically.

\subsection{Overlap and Local energy evaluation}\label{app:ovlploc}
As mentioned above, the overlap and local energy evaluations are the key subroutines
that involve the quantum trial wavefunctions.
One approach to the overlap evaluation is to use the Hadamard
test.\cite{Yu_Kitaev1995-zv} 
Using modern methods, one could do this without requiring the state preparation
circuit to be controlled by an ancilla qubit.\cite{Huggins2020-gi, Lu2020-hw,Russo2021-qx} 
However, this approach would require a separate evaluation for each walker at every
time step. 
To avoid a steep prefactor in quantum device run time, we propose the use of
the technique known as shadow tomography as discussed in
\cref{app:shadow_tomography}.
For now, we will assume that one can make a query to the quantum processor to
obtain the overlap between a quantum trial state and an arbitrary Slater
determinant efficiently up to additive error of the overlap.

With the ability to measure the overlap between $|\Psi_T\rangle$ and an arbitrary single Slater determinant, $|\phi_i(\tau)\rangle$ 
we can easily estimate the local energy in \cref{eq:eloc}.
The evaluation of the denominator is just an overlap quantity and 
an efficient estimation of the denominator is possible via
\begin{equation}
\langle \Psi_T |\hat{H} |\phi_i (\tau)\rangle
=
\sum_{pr} \langle \Psi_T | \phi_{p}^{r} \rangle\langle \phi_{p}^{r}|\hat{H} |\phi_i (\tau)\rangle+
\sum_{pqrs} \langle \Psi_T | \phi_{pq}^{rs} \rangle\langle \phi_{pq}^{rs}|\hat{H} |\phi_i (\tau)\rangle,
\label{eq:eloc2}
\end{equation}
where $| \phi_{p}^{r} \rangle$ and $| \phi_{pq}^{rs} \rangle$
denote 
single and double excitations from $|\phi_i (\tau)\rangle$, respectively.
We only need up to double excitations because our Hamiltonian has up to two-body terms.
It is then evident that
the ability to estimate 
$\langle \Psi_T | \phi_{p}^{r}\rangle$
and
$\langle \Psi_T | \phi_{pq}^{rs}\rangle$ efficiently 
is sufficient to evaluate the entire local energy because
the rest of the terms in \cref{eq:eloc2}
follow from the simple application of the Slater-Condon rule.\cite{Szabo1996Jul}
The number of overlap queries made to the quantum processor
scales as $\mathcal O(N^4)$ with $N$ being the system size in this algorithm.
Other ``mixed'' local observables can be computed via similar algorithms.

\subsection{Virtual correlation energy}\label{app:virtual}
Obtaining the correlation energy outside the ``active'' space, where the actual quantum resource is spent, is critical for converging our simulation results to the basis set limit (or the continuum limit). 
The correlation energy outside the active space will be referred to as ``virtual correlation energy''.
We are limited in terms of the number of qubits on NISQ devices, so a procedure to incorporate correlation energy outside the relatively small active space is essential.
To this end, a virtual correlation energy strategy has been proposed within the framework of VQE,\cite{Takeshita2019} but this approach comes with a significant measurement overhead due to the requirement of three- and four-body reduced density matrices within the active space.

In this section, our goal is to show that a similar technique for QC-AFQMC exists where we can obtain the virtual correlation energy without any additional qubits or any measurement overhead. 
We write our trial wavefunction as
\begin{equation}
|\Psi_T\rangle = |\psi_T\rangle \otimes |\psi_\mathrm{c}\rangle \otimes |0_\mathrm{v}\rangle,
\end{equation}
where $|\psi_T\rangle$ is the quantum trial wavefunction within the active space, $|\psi_c\rangle$ is a Slater determinant 
composed of occupied orbitals outside the active space (i.e. frozen core orbitals), and
$|0_v\rangle$ is a vacuum state in the space of unoccupied orbitals outside the active space (i.e., frozen virtual orbitals).
We want to compute the overlap between $|\Psi_T\rangle$ and a single Slater determinant $|\phi\rangle$
\begin{align}
\braket{\phi}{\Psi_T} = \Bra{\phi} \left(\Ket{\psi_T} \otimes \Ket{\psi_c} \otimes \ket{0_v} \right)
&=
\sum_{\substack{
x \in {\{0, 1\}}^{N_{\mathrm{a}}} \\
y \in {\{0, 1\}}^{N_{\mathrm{c}}} \\
z \in {\{0, 1\}}^{N_{\mathrm{v}}}
}}
\Braket{\phi | x, y, z} 
\Braket{x | \psi_T}
\Braket{y | \psi_c}
\Braket{z | 0_v}
\\
&=
\sum_{\substack{
x \in {\{0, 1\}}^{N_{\mathrm{a}}} \\
y \in {\{0, 1\}}^{N_{\mathrm{c}}} \\
z \in {\{0, 1\}}^{N_{\mathrm{v}}}
}}
\phi^*(x, y, z)
\psi_T(x)
\psi_c(y)
\delta_{z, 0_v}
\\&=
\sum_{
x \in {\{0, 1\}}^{N_{\mathrm{a}}}
}
\left(\sum_{y \in {\{0, 1\}}^{N_{\mathrm{c}}}}
\phi^*(x, y, 0_v) \psi_c(y)
\right)
\psi_T(x),
\end{align}
where $\phi(x, y, z) = \Braket{x, y, z| \phi}$, $\psi_T(x) = \Braket{x | \psi_T}$, $\psi_c(y) = \langle y | \psi_c\rangle$ 
$N_{\mathrm{a}}$ is the number of active spin orbitals, and $N_{\mathrm{c}}$ and $N_{\mathrm{v}}$ are the number of occupied and unoccupied spin orbitals outside of the active space, respectively. 
We are using $x,y,z$ to denote bit strings in the space composed of single particle orbitals used to construct $|\Psi_T\rangle$.
Because the tensor $\phi^*(x, y, z)$ represents a Slater determinant, it is a special case of what is known as a matchgate tensor with $N_{\mathrm{a}} + N_{\mathrm{c}} + N_{\mathrm{v}}$ open indices. This is also the case for $\psi_c(y)$ and $\delta_{z, 0_v}$ (with \(N_\mathrm{c}\) and \(N_\mathrm{v}\) open indices respectively).
Thus, their contraction $
\left(\sum_{y \in {\{0, 1\}}^{N_{\mathrm{c}}}}
\phi^*(x, y, 0_v) \psi_c(y)
\right)$
is also a matchgate with $N_{\mathrm{a}}$ open indices and support on states of a fixed Hamming weight (i.e. an unnormalized Slater determinant), and can be formed efficiently by contracting over $N_{\mathrm{c}} + N_{\mathrm{v}}$ legs with $|\psi_c\rangle \otimes \ket{0_v}$.\cite{bravyi2008contraction,hebenstreit2019all,divincenzo2005fermionic}
Let $\tilde{\phi}(x)$ denote the resulting matchgate tensor after normalization and \(\ket{\tilde{\phi}}\) the associated state. Then \(\ket{\tilde{\phi}}\) is a normalized Slater determinant in the same Hilbert space as \(\ket{\psi_T}\).
Thus, we have
\begin{equation} \braket{\phi}{\Psi_T} = 
\Bra{\phi} \left(\Ket{\psi_T} \otimes \Ket{\psi_c} \otimes \ket{0_v} \right)
=
\text{constant} \times \Braket{\tilde{\phi} | \psi_T},
\label{eq:matchgateovlp}
\end{equation}
where the constant can be efficiently evaluated classically by contracting matchgate states
and
the evaluation of $\Braket{\tilde{\phi} | \psi_T}$ can now be performed on the quantum computer with only $N_{\mathrm{a}}$ qubits.

For the local energy evaluation in \cref{eq:eloc}, we leverage the same technique that we used in \cref{eq:eloc2}.
The numerator of the local energy expression is
\begin{align}
\langle\phi|\hat{H}|\left(\Ket{\psi_T} \otimes \Ket{\psi_c}\otimes|0_v\rangle\right)
&=
\sum_{pr}\langle\phi|\hat{H}|\phi_{p}^{r}\rangle\langle\phi_{p}^{r}| \left(\Ket{\psi_T} \otimes \Ket{\psi_c} \otimes \ket{0_v} \right)+
\sum_{pqrs}\langle\phi|\hat{H}|\phi_{pq}^{rs}\rangle\langle\phi_{pq}^{rs}| \left(\Ket{\psi_T} \otimes \Ket{\psi_c} \otimes\ket{0_v}\right),
\end{align}
and we only need to focus on the computing the following term:
\begin{equation}
\langle\phi_{pq}^{rs}| \left(\Ket{\psi_T} \otimes \Ket{\psi_c} \otimes \ket{0_v}\right)
=
\sum_{\substack{
x \in {\{0, 1\}}^{N_{\mathrm{a}}} \\
y \in {\{0, 1\}}^{N_{\mathrm{c}}}
}}
\phi_{pq}^{rs}(x,y, 0_v)\psi_T(x) \psi_c(y)
=
\sum_{\substack{
x \in {\{0, 1\}}^{N_{\mathrm{a}}}
}}
\left(\sum_{y \in {\{0, 1\}}^{N_{\mathrm{c}}}}\phi_{pq}^{rs}(x,y,0_v)\psi_c(y)\right)\psi_T(x).
\end{equation}
Then $\left(\sum_{y \in {\{0, 1\}}^{N_{\mathrm{c}}}}\phi_{pq}^{rs}(x,y,0_v)\psi_c(y)\right)$ is the tensor corresponding to a matchgate state itself (with $N_{\mathrm{a}}$ open indices) and thus can be computed efficiently classically. 
Since an equation of the form \cref{eq:matchgateovlp} also holds for $|\phi_{pq}^{rs}\rangle$, the local energy evaluation can be performed on the quantum computer with only $N_{\mathrm{a}}$ qubits.

\section{Experimental Implementation via Shadow Tomography}
\label{app:shadow_tomography}

The basic goal of shadow tomography is to estimate properties of a quantum
state without resorting to full state tomography.
This task was introduced in \citen{Aaronson2020-ei} and has been considered in a
number of subsequent works.\cite{Huang2020-us, Chen2020-cq, Struchalin2020-wm,
  Koh2020-fw, Zhao2020-fv, Aharonov2021-yu, Huang2021-ed, Hadfield2021-wm,
  Hu2021-of}
In the experiments performed in this work, we make use of these tools to
approximate the quantities required to perform AFQMC, \cref{eq:energy} and
\cref{eq:ovl}.
We focus here on the proposal put forward by Huang et al. 
in \citen{Huang2020-us}. 
This version of shadow tomography is experimentally simple to implement and
compatible with today's quantum hardware. 

As we shall explain, the use of shadow tomography makes our experiment
particularly efficient in terms of the number of repetitions required to
evaluate the required wavefunction overlaps.
This allows us to avoid performing a separate set of experiments (e.g. 
using the Hadamard test) for each timestep and walker. 
However, this efficiency comes at a cost; the way in which we extract these
overlaps from the experimental measurement record requires an exponentially
scaling post-processing step.
We note that this difficulty is specific to the particular choice we made to demonstrate QC-QMC using AFQMC rather than some other QMC method. For example, if we were using a quantum computer to provide the constraint for a Green's function Monte Carlo calculation, the walker wavefunctions would be computational basis states and we could make use of shadow tomography without this issue.
It is an open question whether a more sophisticated measurement strategy could be
equally efficient in terms of the number of measurements required while also
avoiding this additional bottleneck for QC-AFQMC. Exploring the use of shadow tomography with
random fermionic gaussian circuits, as in \citen{Zhao2020-fv}, seems like a promising direction to explore for this purpose.

In \app{shadow_tomo_review}, we review the general formalism of shadow
tomography as proposed in \citen{Huang2020-us}.
We continue in \app{shadow_tomo_overlap} by showing how we can use shadow
tomography to approximate the wavefunction overlaps required to perform QC-QMC
and discussing the scaling in terms of the number of measurement repetitions
performed on the quantum device. 
We explain the challenges associated with the classical post-processing of the
experimental record for QC-AFQMC in \app{vmc}. 
In \app{global-stabilizer-measurements} and \app{partitioned_shadow_tomo}, we
describe two strategies we adopt for reducing the number of quantum gates
required for our experimental implementation.
\app{global-stabilizer-measurements} deals with compiling the measurements,
while \app{partitioned_shadow_tomo} explains how we make a tradeoff between the
number of gates and the number of measurements.
Finally, in \app{noise_resilience}, we show that the quantities we ultimately
estimate using the quantum device are resilient to noise, particularly noise
during the shadow tomography measurement procedure.

\subsection{Review of Shadow Tomography}
\label{app:shadow_tomo_review}

Let \(\rho\) denote some unknown quantum state. 
We assume that we have access to \(N\) copies of \(\rho\). 
Let \(\{O_i\}\) denote a collection of \(M\) observables. 
Our task is to estimate the quantities \(\tr(\rho O_i)\) up to some additive
error \(\epsilon\) for each \(O_i\). 
The key insight of \citen{Huang2020-us} is that we can accomplish this
efficiently in certain circumstances by randomly choosing measurement operators
from a tomographically complete set.

To specify a protocol, we choose an ensemble of unitaries \(\mathcal{U}\). 
We then proceed by randomly sampling \(U_k \in \mathcal{U}\) and measuring the
state \(U_k \rho U_k^\dagger\) in the computational basis to obtain the basis
state \(\ketbra{b_k}\).
Consider the state \(U_k^\dagger \ketbra{b_k} U_k\). 
In expectation, the mapping from \(\rho\) to \(U_k^\dagger \ketbra{b_k} U_k\) defines a quantum
channel,
\begin{equation}
  \mathcal{M}(\rho) \coloneqq \mathbb{E}_k\big[ U_k^\dagger \ketbra{b_k} U_k \big].
  \label{eq:shadow_channel_def}
\end{equation}
We require that \(\mathcal{M}\) be invertible, which is true if and only if the
collection of measurement operators defined by drawing \(U \in \mathcal{U}\) and
measuring in the computational basis is tomographically complete.
Assuming that this is true, we can apply \(\mathcal{M}^{-1}\) to both
sides of \eq{shadow_channel_def}, yielding
\begin{align}
  \label{eq:shadow_definition}
  \rho =& \mathcal{M}^{-1}\Big(\mathbb{E}_k\big[ U_k^\dagger \ketbra{b_k} U_k
          \big]\Big) \nonumber \\
  =& \mathbb{E}_k\Big[ \mathcal{M}^{-1}\big( U_k^\dagger \ketbra{b_k} U_k \big)\Big].
\end{align}
We call the collection \( \Big\{ \mathcal{M}^{-1}\big( U_k^\dagger \ketbra{b_k}
U_k \big) \Big\} \) the classical shadow of \(\rho\).

Many choices for the ensemble \(\mathcal{U}\) are possible.\cite{Huang2020-us,Zhao2020-fv, Hadfield2021-wm, Hu2021-of}
Formally, the condition that the measurement channel is invertible is
sufficient.
In practice, it is also desirable to impose the constraint that the classical
post-processing involved in making use of the shadow can be done efficiently.
In this work, we utilize randomly selected $N$-qubit Clifford circuits, as well
as tensor products of randomly selected Clifford circuits on fewer qubits.

\subsection{Approximating Wavefunction Overlaps with Shadow Tomography}
\label{app:shadow_tomo_overlap}
Let \(\ket{\Psi_T}\) denote our trial wavefunction. 
We restrict ourselves to considering \(\ket{\Psi_T}\) that represent fermionic
wavefunctions with a definite number of particles $\eta > 0$.
We focus on states encoded with the Jordan-Wigner transformation, so that the
qubit wavefunction for \(\ket{\Psi_T}\) is a superposition of computational
basis states with Hamming weight \(\eta\).
Let \(\ket{\phi}\) denote our walker wavefunction, which is also a
superposition of computational basis states with Hamming weight \(\eta\).
In this section, we explain how to approximate the wavefunction overlap
$\braket{\phi}{\Psi_T}$ using shadow tomography.

Our protocol begins by preparing the state $\ketbra{\tau}{\tau}$ on the quantum
computer, where $\Ket{\tau} = (\Ket{0} + \Ket{\Psi_T}) / \sqrt{2}$, with
\(\ket{0}\) denoting the all-zero (vacuum) state. 
The wavefunction overlap of interest is therefore equal to
\begin{equation}\label{eq:overlap_basis_state}
\braket{\phi}{\Psi_T} = 2 \langle \phi \ketbra{\tau} 0 \rangle = 
2 \Tr\left[\ketbra{\tau}\cdot \ketbra{0}{\phi} \right],
\end{equation}
where we used the fact that $\Braket{\Psi_T | 0} = \Braket{\phi|0} = 0$.
If we define the observables
\begin{align}
  P_{+} &= \ketbra{0}{\phi} + \ketbra{\phi}{0}, \nonumber \\
  P_{-} &= -i(\ketbra{0}{\phi} -\ketbra{\phi}{0}),
\end{align}
then we have
\begin{align}
\Re(\braket{\phi}{\Psi_T}) &= \Tr\left[\Ket{\tau}\Bra{\tau}P_+\right],\\
\Im(\braket{\phi}{\Psi_T}) &= \Tr\left[\Ket{\tau}\Bra{\tau}P_-\right],
\end{align}
where $z = \Re(z) + i \Im(z)$ for $z \in \mathbb{C}$.
Note that $\Tr\left[P_{\pm}\right]=0$ and 
\begin{equation}
\Tr\left[P_{\pm}^2\right]=
\Tr\left[
\ketbra{\phi} + \ketbra{0}
\right]
= 2.
\end{equation}
assuming $|\phi\rangle$ is a normalized wavefunction.

Let us assume for now that \(\mathcal{U}\) is the Clifford group on $N$ qubits.
Therefore, we can use the expression for the inverse channel from \citen{Huang2020-us},
\begin{equation}
  \label{eq:channel_inverse_n_clifford}
  \mathcal{M}^{-1} \big(X) = (2^N + 1)X - \mathbb{I},
\end{equation}
where \(X\) is a placeholder variable.
In particular, we have
\begin{align}
  \label{eq:channel_inverse_particular}
  \Tr \big[ (P_+ + i P_-) \ketbra{\tau} \big] \approx 
  \Tr \big[ (P_+ + iP_-) \mathcal{M}^{-1}\big( U_k^\dagger \ketbra{b_k} U_k \big) &\big] 
  = 
  (2^N + 1) \Tr \big[ (P_+ + iP_-) U_k^\dagger \ketbra{b_k} U_k \big].
\end{align}
The full expression for \(\braket{\phi}{\Psi_T}\) then
becomes
\begin{align}
  \label{eq:overlap_shadow_estimator}
  \braket{\phi}{\Psi_T} = (2^N + 1) \mathbb{E}_k \Big[ \Tr \big[ (P_+ + i P_-) U_k^\dagger \ketbra{b_k} U_k \big] \Big] = \\
  2(2^N + 1) \mathbb{E}_k \Big[  \bra{\phi} U_k^\dagger \ketbra{b_k} U_k \ket{0}  \Big].
  \label{eq:shadow_tomo_estimator}
\end{align}

Furthermore, because we are expressing \(\braket{\phi}{\Psi_T}\) in terms of the expectation values of the two operators \(P_\pm\) with $\Tr[P_{\pm}^2]=O(1)$, Theorem 1 of \citen{Huang2020-us} allows us to bound the number of measurement
repetitions we require for a target precision. Specifically, when we take the ensemble of random unitaries to be the Clifford group on all \(N\) qubits, as we do in this section, this bound scales with the Hilbert-Schmidt norm of the operators of interest. 
Consider the case where we would like to estimate the overlap of
\(\ket{\Psi_T}\) with a collection of \(M\) different wavefunctions \(\{\phi_i\}\).
Let \(\tilde{c_i}\) denote our estimate of \(\braket{\phi_i}{\Psi_T}\). 
We specify a desired accuracy in terms of two parameters, \(\epsilon\) and
\(\delta\), by demanding that
\begin{equation}
  \label{eq:epsilon_acc_def}
  |\tilde{c_i} - \braket{\phi_i}{\Psi_T}| \leq \epsilon \;\; \forall \;\; 1 \leq i \leq M
\end{equation}
with probability at least \(1 - \delta\). 
Theorem 1 of \citen{Huang2020-us} implies that shadow tomography using the
$N$-qubit Clifford group allows us to achieve this accuracy using
\begin{equation}
  \label{eq:shadow_tomo_reps}
  R = \mathcal{O}\big( \frac{\log(M) - \log(\delta)}{\epsilon^2} \big)
\end{equation}
repetitions of state preparation and measurement.

\subsection{Classical Post-processing for Wavefunction Overlaps}\label{app:vmc}
In the previous section, we described how we can use shadow tomography to
estimate overlaps of the form $\Braket{\phi | \Psi_T}$ by evaluating the
expression in \eq{shadow_tomo_estimator}, \(2(2^N + 1) \mathbb{E}_k \Big[ \bra{\phi} U_k^\dagger \ketbra{b_k}
U_k \ket{0} \Big]\), where the $U_k$ are Clifford circuits and $b_k$ are
computational basis states. 
We explained how these estimates can be made using a modest number of
experimental repetitions, even for a large collection of different
\(\ket{\phi_i}\). However, we have not yet described the classical
post-processing required to perform this estimation. This section addresses this
aspect of our experiment and explains how the approach we took for our implementation of QC-AFQMC in practice involves an
exponentially scaling step.
We will utilize the fact that overlap between stabilizer states (including basis
states) can be efficiently computed classically using the Gottesman-Knill
theorem.\cite{Gottesman1996-yg,Aaronson2004-yb}
For instance, the terms \(\mel{b_k}{U_k}{0}\) can be efficiently calculated for
any Clifford circuit \(U_k\).
Therefore, we can just focus on computing \(\mel{\phi}{U_k^\dagger}{b_k}\) to evaluate the expression in \eq{shadow_tomo_estimator}.

In special cases, this can be computed efficiently.
For example, if $\Ket{\phi} = \sum_{\alpha} c_{\alpha} \Ket{\phi_{\alpha}}$ can be written as
a linear combination of a polynomial number of stabilizer states
${\left\{\Ket{\phi_{\alpha}}\right\}}_{\alpha}$, then we can efficiently compute
$\mel{\phi_{\alpha}}{U^\dagger_k}{b_k}$ for each $\alpha$ and sum them together. QMC methods such as Green's function Monte Carlo where the walker wavefunctions are computational basis states are a special case that trivially satisfies this requirement.
Even when $\Ket{\phi}$ is not exactly sparse, it may be approximately sparse in
the computational basis (in the sense of being close to an exactly sparse
state). 
In such a case, provided that we can sample from \(\ket{\phi}\) efficiently (which is
possible for a Slater determinant), we could construct a sparse approximation to \(\ket{\phi}\) (see, e.g., \citen{schwarz2013simulating}) and use this state to approximate the overlap.
In our QC-AFQMC experiments, we expanded \(\ket{\phi}\) in this way, except that we performed a
sum over all of the computational basis states with the correct symmetries,
incurring an exponential overhead.
We emphasize, however, that the cost of this post-processing has no effect on
the number of quantum samples needed to produce the classical shadow.

For a general wavefunction $\Ket{\phi}$, computing $\Braket{\phi|U_k^\dagger |b_k}$ may be classically intractable. 
Specifically, when $\Ket{\phi}$ is a Slater determinant, as our walkers are,
there is no known way to efficiently
compute 
the desired overlap classically.
Existing strategies for approximating the overlap between
two states can allow us to bypass this exponential scaling if an additive error is acceptable. In general, it is possible to approximate the overlap between two states up to
some additive error provided that one can sample from one of the states in the
computational basis and query each of them for the amplitudes of particular
bitstrings.
Techniques of this sort are used in variational Monte Carlo\cite{Becca2017Nov} and have also been studied in the context of
dequantizing quantum algorithms. 
In particular, \citen{tang2019quantum} showed that
for normalized states $\Ket{\psi}$, $\Ket{\phi}$, the random
variable $\frac{\Braket{\phi|x}}{\Braket{\psi|x}}$ with probability
${\left|\Braket{x|\psi}\right|}^2$ has mean $\Braket{\psi|\phi}$ and constant
variance:
\begin{equation}\label{eq:ovlmet}
  \langle
  \psi | \phi
  \rangle
  =
  \sum_x
  \langle
  \psi \ketbra{x} \phi
  \rangle
  =
  \sum_x
  \frac{\langle
  \psi |x \rangle }
  {
  \langle \phi | x \rangle
}
  |\langle x | \phi
  \rangle|^2
  .
\end{equation}
This implies an algorithm to calculate $\Braket{\psi|\phi}$ to within $\epsilon$
additive error with failure probability at most $\delta$ using
$O(\frac{1}{\epsilon^2} \log \frac{1}{\delta})$ samples from $\Ket{\psi}$ and
queries to the amplitudes of $\Ket{\psi}$ and \(\ket{\phi}\). 
Unfortunately, the prefactor of \(2(2^N + 1)\) in \eq{shadow_tomo_estimator}
seems to preclude benefiting from a strategy that estimates
\(\mel{\phi}{U_k^\dagger}{b_k}\) up to an additive error. This is why we chose to compute the overlap using the exponential scaling enumeration of basis states in our QC-AFQMC experiments.

\subsection{Global Stabilizer Measurements}
\label{app:global-stabilizer-measurements}

In this section, we outline a strategy for reducing the size of the circuits required to perform shadow tomography. This strategy leverages the fact that we measure in the computational basis immediately after performing a randomly sampled Clifford. Therefore, any permutation of the computational basis states that occurs immediately prior to measurement is unnecessary. 

In general, applying a unitary $U$ and then measuring in the computational basis
$\left\{ \ket{\mathbf x}: \mathbf x \in {\{0, 1\}}^N\right\}$, as shadow
tomography was originally presented, is equivalent to measuring in the rotated
basis $\left\{U^{\dagger} \ket{\mathbf x} : \mathbf x \in {\{0, 1\}}^N\right\}$.
For a set of unitaries $\mathcal U$, choosing a unitary therefrom uniformly at
random and then measuring in the computational basis is equivalent to measuring
the positive operator-valued measure (POVM) $\left\{ \frac{1}{ \left|\mathcal U\right| } U^{\dagger} \ket{\mathbf
    x} \bra{\mathbf x} U : \mathbf x \in {\{0, 1\}}^N, U \in \mathcal U \right\}
$ .
Note that the $\left|\mathcal U\right| 2^N$ measurement operators need not be
distinct (e.g., if the unitaries in $\mathcal U$ only permute the computational
basis states).
In particular, when $\mathcal U$ is the set of $N$-qubit Clifford unitaries
$\mathcal C_N$, each measurement operator $U^{\dagger} \ket{\mathbf x}
\bra{\mathbf x} U$ is a stabilizer state, and the POVM is
\begin{equation}
  \left\{ \frac{2^N}{\left|\stab_N\right|} \ket{\psi} \bra{\psi} : \ket{\psi} \in \stab_N\right\},
\end{equation}
where $\stab_N$ is the set of $N$-qubit stabilizer states.
That the weight of the measurement operators is uniform follows from the
symmetry of $\mathcal U$ (appending any Clifford to each $U \in \mathcal U$
leaves the distribution unchanged); that the uniform weight is $2^N /
\left|\stab_N\right|$ will be explained later.
There are $\left|\mathcal C_N\right| = 2^{N^2 + 2N} \prod_{i=1}^N (4^i - 1)$
Clifford unitaries~\cite{bravyi2020hadamardfree} and only ${2^N \prod_{i=1}^N
  (2^i + 1)} \ll 2^N \left|\mathcal C_N\right| $ stabilizer
states.\cite{Aaronson2004-yb}
This suggests that sampling a uniformly random Clifford is unnecessary.
We will now construct a smaller set of $2^{-n} \left|\stab_N\right|$ unitaries
$\tilde{\mathcal C}_N$ such that the corresponding POVM is equivalent to that of
$\mathcal C_N$.
Specifically, $\stab_N = \left\{ U^{\dagger} \ket{\mathbf x} : U \in
  \tilde{\mathcal C}_N, \mathbf x \in {\{0, 1\}}^N\right\}$.

Let $\mathcal F_N$ be the ``H-free'' (Hadamard-free) group on $N$ qubits, i.e. 
the group generated by X, CNOT, CZ.
The action of any H-free operator can be written
as~\cite{bravyi2020hadamardfree}
\begin{equation}
  F(\Gamma, \boldsymbol \gamma, \Delta, \boldsymbol \delta) \ket{\mathbf x}
  =
  i^{\mathbf x^{\transpose} \Gamma \mathbf x} {(-1)}^{\boldsymbol \gamma \cdot \mathbf x} \ket{\Delta \mathbf x + \boldsymbol \delta}
  ,
\end{equation}
where $\Gamma$ is symmetric Boolean matrix; $\boldsymbol \gamma, \boldsymbol \delta
\in {\{0, 1\}}^N$; and $\Delta$ is an invertible Boolean matrix.
(A Boolean matrix is one whose entries are 0 or 1.)
The action of an H-free operator thus is to simply permute the basis states and
add some phase.
If we are measuring in the computational basis, the phase does not affect the
outcome probabilities and the affine change $\mathbf x \mapsto \Delta \mathbf x
+ \boldsymbol \delta$ is invertible.
Therefore measuring a state in the computational basis and applying the
transformation $\mathbf y \mapsto \Delta^{-1} (\mathbf y + \boldsymbol \delta)$
to the outcome $\mathbf y$ is equivalent to applying $F^{\dagger}$ and then measuring in
the computational basis (i.e., measuring in the basis $\left\{F \ket{\mathbf x}: \mathbf x \in {\{0, 1\}}^N\right\}$).
As shown by Bravyi and Maslov,\cite{bravyi2020hadamardfree} any Clifford operator can be written in the form
$F \cdot H \cdot F'$, where $F, F' \in \mathcal F_N$ and $H$ is a layer of
single-qubit Hadamards.
In shadow tomography, we apply a Clifford $F \cdot H \cdot F'$ and measure in
the computational basis.
As explained above, however, the second H-free operator $F$ need not actually be
applied; its effect can be implemented entirely in classical post-processing.
In general, $F$ and $F'$ are not unique.
However, Bravyi and Maslov give a canonical form for Clifford operators (by
constraining the H-free operators $F, F'$) that allows for uniform sampling.
If we start with their canonical form and ``push'' as much of $F'$ through the
Hadamard layer into $F$, yielding a new form $\tilde{F} \cdot H \cdot \tilde{F}'
= F \cdot H \cdot F'$, and neglect the new final H-free operator $\tilde{F}$, we
are left with an operator of the form
\begin{equation}
  G(I, \Gamma, \Delta)
  =
  \prod_{i \in I} H_i P_i^{\Gamma_{i, i}}
  \prod_{\substack{i \in I \\ j \in I: j \neq i}} \CZ_{i, j}^{\Gamma_{i, j}}
  \prod_{\substack{
      i \in I\\
      j \notin I: j > i
    }
  } \CX_{i, j}^{\Delta_{i, j}}
  ,
  \label{eq:truncated-clifford}
\end{equation}
where $I \subset [N]$ is a subset of qubit indices, $\Gamma$ is a Boolean
upper-triangular matrix with support only on $I$, and $\Delta$ is Boolean.
Applying a Clifford operator and measuring in the computational basis can thus
be replaced by applying an operator of the form in~\eq{truncated-clifford} and
measuring in the computational basis. 
A priori, we would also need to do post-processing to account for the affine
transformation effected by the neglected H-free operator, but in fact this is
not needed.


\subsection{Partitioned Shadow Tomography}
\label{app:partitioned_shadow_tomo}
As we discussed in \app{shadow_tomo_overlap}, shadow tomography using the
$N$-qubit Clifford group can be used to simultaneously estimate \(M\) wavefunction
overlaps using a number of samples that scales logarithmically in \(M\). 
However, performing these measurements on a NISQ devices can be challenging
because of the required circuit depth. 
Alternative choices of the ensemble of random unitaries, \(\mathcal{U}\), can
alleviate this difficulty. 
In \citen{Huang2020-us}, Huang et al.\ consider a second choice of \(\mathcal{U}\)
where the unitaries \(U \in \mathcal{U}\) are instead chosen to be tensor
products of single-qubit Clifford operators. This choice leads to especially
simple circuits. In the worst case, however, it requires a number of
measurements scaling exponentially with the locality of the operators to be
estimated.

In the experiments performed in this work, we found it useful to interpolate
between these two extremes. Specifically, we use an ensemble of random circuits
\(\mathcal{U}\) consisting of tensor products of random Clifford circuits on \(N
/ 2\) qubits. In this section, we explain how the the techniques for overlap
estimation we presented in \app{shadow_tomo_overlap} can be generalized to this
case. \citen{Huang2020-us} explains how each choice of \(\mathcal{U}\) has an
associated norm which can be used to bound the variance of the estimators
derived from the classical shadow. We do not work out the norm or the associated bounds on the number of measurements for our
partitioned shadow tomography here. Instead, we merely note that it performed
well in practice and leave this elaboration for a future work.

Recalling and simplifying the expression in \eq{channel_inverse_particular}, we have
\begin{equation}
  \label{eq:part_shadow_key}
  \braket{\phi}{\Psi_T} = 2 \mathbb{E}_k\Big[ \bra{\phi} \mathcal{M}^{-1}\big( U_k^\dagger \ketbra{b_k} U_k \big) \ket{0} \Big].
\end{equation}
We can use an expression like the one from \eq{channel_inverse_n_clifford} to
apply the inverse channel, but first we need to specify some notation.
We take a partitioning of the \(N\) qubits into \(P\) parts. 
Let \(N_1, N_2,... 
N_P\) be the number of qubits in each part of the partition. 
We consider a shadow tomography protocol that applies a randomly selected
\(N_p\)-qubit Clifford to each part, \(p \in \{1, 2, ... 
P\}\). 
Thus, we have
\begin{equation}
  \label{eq:partitioned_unitary}
  U_k = U_k^{1} \otimes U^{2}_k \otimes \cdots U_k^{P}.
\end{equation}
The inverse of the shadow tomography measurement channel is simply
\begin{equation}
  \label{eq:partitioned_M_inv}
  \mathcal{M}^{-1} = \bigotimes_{p=1}^P \mathcal{M}^{-1}_{N_p},
\end{equation}
where, as in \eq{channel_inverse_n_clifford},
\begin{equation}
  \label{eq:part_M_inv}
  \mathcal{M}^{-1}_{N_p}\left( X\right) = (2^{N_p} + 1) X - \mathbb{I}_{N_p}.
\end{equation}
where $X$ is a placeholder variable.

Now we specialize to the case where \(\ket{\phi}\) is a computational basis
state, which we denote by \(\ket{\beta}\). 
We could instead take \(\ket{\phi}\) to be any state which is separable
between the parts of the partition (or a sum of such states), but specializing
to computational basis states is sufficient for our purposes.
Let \(\ket{\beta_p}\) denote the component of \(\ket{\beta}\) associated with
the \(p\)-th part of the partition.
Using this notation, we can evaluate \eq{part_shadow_key} to yield
\begin{equation}
  \label{eq:partitioned_overlap_simplified}
  \braket{\beta}{\Psi_T} = 
   2 \mathbb{E}_k \Big[ \prod_{p=1}^P (2^{N_p} + 1)\bra{\beta_p}U_k^{p \dagger} \ketbra{b_k^p} U_k^p \ket{0_p} - \braket{\beta_p}{0_p}  \Big].
\end{equation}

In carrying out our experiments, we specifically chose to use a partition with
two parts, one for each of the spin sectors. 
All of our walker wavefunctions \(\ket{\phi}\) were superpositions of basis
states with a Hamming weight \(\eta\) overall and a nonzero number of electrons in each spin sector. 
Therefore, when we used shadow tomography to evaluate the overlap of our walker
wavefunctions \(\ket{\phi}\) with \(\ket{\Psi_T}\) as described in
\app{shadow_tomo_overlap} and \app{vmc}, \(\braket{\beta_p}{0_p} = 0\) for the calculations we performed. Because of this, we were able to evaluate the wavefunction overlaps using the expression
\begin{equation}
  \label{eq:overall_partitioned_overlap}
  \braket{\phi}{\Psi_T} = \sum_i c_i \braket{\beta^i}{\Psi_T} = 
  \sum_{i} c_i
  2 \mathbb{E}_k \Big[ \prod_{p=1}^P (2^{N_p} + 1)\bra{\beta^{i}_p}U_k^{p \dagger} \ketbra{b_k^p} U_k^p \ket{0_p} \Big],
\end{equation}
where the \(c_i\)'s are the amplitudes of \(\ket{\phi}\) in the computational basis, $\{|\beta^i\rangle\}$.


\subsection{Noise Resilience}
\label{app:noise_resilience}

We show in this section that, in certain circumstances, noise has a negligible impact on the measurement of overlap
ratios such as
\begin{equation}\label{eq:overlap_ratio_noise_sec}
  \frac{\Braket{\phi_1|\Psi_T}}{\Braket{\phi_2|\Psi_T}},
\end{equation}
where \(\ket{\Psi_T}\) is some fixed trial wavefunction and \(\ket{\phi_1},
\ket{\phi_2}\) are two arbitrary determinants.
Recall that the overlap $\Braket{\phi_i | \Psi_T} = 2 \Braket{\phi_i | \rho | 0}$, where $\rho = |\tau\rangle\langle \tau|= (\ket{0} + \ket{\Psi_T})(\Bra{0}+\Bra{\Psi_T}) /2$.

As a warm up, consider a simple noise model: a global depolarizing channel\cite{Nielsen2010Dec}
\begin{equation}
\rho \mapsto \rho' = (1 - p) \rho + p \mathbb{I}
\end{equation}
applied right before measurement.
Then, neglecting the error in estimating the overlaps due to measurement, our estimate of the overlap becomes
\begin{align}
\frac{2\Braket{\phi_1|\rho'|0}}{2\Braket{\phi_2|\rho' | 0}}
&=
\frac{\Braket{\phi_1|\rho'| 0}}{\Braket{\phi_2|\rho' | 0}}
\\
&=
\frac{
(1-p)\Braket{\phi_1|\rho| 0} + p \Braket{\phi_1|0}
}{
(1-p)\Braket{\phi_2|\rho | 0} + p \Braket{\phi_2|0}}
\\
&=
\frac{(1-p)\Braket{\phi_1|\rho| 0}}{(1-p)\Braket{\phi_2|\rho | 0}}
\\
&=
\frac{\Braket{\phi_1|\rho| 0}}{\Braket{\phi_2|\rho | 0}}
,
\end{align}
where we used the fact that $\Braket{\phi_i | 0} = 0$.
Thus the depolarizing channel has no effect on our estimate.


Now suppose we were to apply the robust shadow tomography procedure of \citen{Chen2020-cq} to determine the overlap ratio in \eq{overlap_ratio_noise_sec}. We will assume for now that the state \(\rho\) is prepared without error and that we have some unknown noise process occurring during the shadow tomography procedure. We focus first on the case where our ensemble of random unitaries (\(\mathcal{U}\)) is the Clifford group on all $N$ qubits, which we refer to as the global case.
First, we would estimate a noise parameter $f$.
Then we would calculate the classical shadow using the inverse channel 
\begin{equation}
\mathcal{M}^{-1} (X) = 
f^{-1} X - 
\frac{1 - f^{-1}}{2^N}  \mathbb{I}
,
\end{equation}
where X is a placeholder variable. Note that, in the absence of noise, we have \(f^{-1} = 2^N + 1\) and we recover \eq{channel_inverse_n_clifford}.
This yields a single-round estimate of the overlap, 
\begin{align}
2 \Braket{\phi_i | \mathcal{M}^{-1}(U^\dagger_k\Ket{b_k}\Bra{b_k}U_k) | 0}
&=
2 f^{-1} \Braket{\phi_i | U^\dagger_k | b_k}\Braket{b_k| U_k | 0}
-
\frac{1 - f^{-1}}{2^N} \Tr[U^\dagger_k \Ket{b_k} \Bra{b_k} U_k] \Braket{\phi_i | 0}
\\
&=
2 f^{-1} \Braket{\phi_i | U^\dagger_k | b_k}\Braket{b_k| U_k | 0}
.
\end{align}
As above, the factor of $f^{-1}$ drops out when taking ratios.
Therefore, when doing shadow tomography (using global Cliffords) to calculate ratios as above, we get robustness \emph{for free}.
That is, we can use the true value in the noiseless case $f={(2^N + 1)}^{-1}$ as in vanilla shadow tomography and the estimates for the ratios are exactly the same as if we had done robust shadow tomography, \emph{without actually doing robust shadow tomography} (i.e., estimating $f$ and using that estimate to obtain the corrected inverse channel).
This is true whenever the assumptions of robust shadow tomography hold, i.e., that the noise is gate-independent, time-stationary and Markovian. 

For partitioned shadow tomography with two partitions (as described in \app{partitioned_shadow_tomo}), the same conclusion holds. \citen{Chen2020-cq} describes in detail how robust shadow tomography applies to to a random ensemble consisting of a tensor product of single-qubit Clifford operators. We can apply the same logic to the case when we have a tensor product of random \(\frac{N}{2}\)-qubit Cliffords. This yields an inverse channel,
\begin{subequations}
\begin{align}
\mathcal{M}^{-1}(\rho)
=
&
\Bigg[
2^{-n}
\left(
f_{0,0}^{-1} - f_{0, 1}^{-1} - f_{1, 0}^{-1} + f_{1, 1}^{-1}
\right)
\mathbb{I}_n
\label{eq:bipartitioned-inverse-channel}
\\
&\quad
+
2^{-n/2}
\left(
f_{0,1}^{-1} - f_{1, 1}^{-1}
\right)
\left(
\mathbb{I}_{n/2} \otimes
\Tr_{P_1} \left[\rho\right]
\right)
\\
&\quad
+
2^{-n/2}
\left(
f_{1,0}^{-1} - f_{1, 1}^{-1}
\right)
\left(
\Tr_{P_2} \left[\rho\right]
\otimes
\mathbb{I}_{n/2} 
\right)
\\
&\quad
+
f_{1, 1} \rho
\Bigg]
\label{eq:bipartitioned-inverse-channel-last}
,
  \end{align}
\end{subequations}
where \(f_{0,0}, f_{0,1}, f_{1,0}, f_{1,1}\) are four four parameters which characterize the impact of the noise. These parameters could be learned from calibration experiments, but, as we will see, this is unnecessary for our purposes.

In our particular case, the two partitions correspond to two spin sectors.
We will assume that $\ket{\psi_i}$ has no overlap with any state of the form $\ket{0} \otimes \ket{\psi}$ or $\ket{\psi} \otimes \ket{0}$; in other words, that the state always has at least one particle of each spin.
Now again consider a single-round estimate of the overlap
$
2 \Braket{\phi_i | \mathcal{M}^{-1}(U^\dagger_k\Ket{b}\Bra{b}U_k) | 0}
$,
where $U_k = U_1 \otimes U_2$ and $\ket{b_k} = \ket{b_1} \otimes \ket{b_2}$.
There will be four contributions, corresponding to \eq{bipartitioned-inverse-channel}--\eq{bipartitioned-inverse-channel-last}.
The first is zero because $\Braket{\phi_i | 0} = 0$.
The second is proportional to 
\begin{align}
\Braket{\phi_i | \mathbb I_{n/2} \otimes \Tr_{P_1} [U^\dagger_k \ket{b} \bra{b} U_k] | 0} &=  
\Braket{\phi_i | \mathbb I_{n/2} \otimes U_2^\dagger \ket{b_2} \bra{b_2} U_2] | 0} \\
& 
\propto 
\Bra{\phi_i} \left(\Ket{0} \otimes U^\dagger_2 \Ket{b_2}\right) = 0.
\end{align}
The third is also zero for the same reason.
That leaves just the last term, so that 
\begin{equation}
2 \Braket{\phi_i | \mathcal{M}^{-1}(U_k^\dagger\Ket{b}\Bra{b}U_k) | 0}
=
2^{-n} f_{1, 1}
\Braket{\phi_i | U^\dagger_k | b}  \Braket{b | U_k | 0}
.
\end{equation}

Therefore, the inverse channel for the noisy implementation of this form of partitioned shadow tomography would simply rescale all of the estimated overlaps by the same noise parameter (when compared with the inverse channel in the absence of noise). This rescaling cancels out when we calculate the overlap ratios and we get robustness automatically whenever the assumptions of robust shadow tomography are satisfied, just as in the global case.


\section{Computational and Experimental Details and Supportive Numerical Results}\label{app:comp}
We used quantum computing tools provided in Cirq,~\cite{developerscirq} 
qsim,~\cite{quantum_ai_team_and_collaborators_2020_4023103} 
and Fermionic Quantum Emulator.~\cite{fqe2021}
For the shadow tomography experiment, we executed each Clifford circuit measurement 1000 times.

All AFQMC calculations presented here were performed with
PAUXY~\cite{pauxy} and QMCPACK.~\cite{Kent2020May}
All integrals are obtained using PySCF \cite{PYSCF} and some of the calculations were verified using Q-Chem.\cite{Shao2015}
Exact energies within a basis were all obtained using a brute-force approach called heat-bath configuration interaction (HCI).\cite{Holmes2016Aug}
For AFQMC, we used more than 1000 walkers in all numerical data presented here to ensure that the population control bias is negligible. $\Delta t = 0.005$ was used for the time step and the resulting time step error was found to be insignificant.
When choosing a set of orbitals for the active space, we decided to use orbitals from a brute-force complete active space self-consistent field (CASSCF) calculation.
This is not really a necessary component in our method and in the future one may determine those orbitals by performing an active-space self-consistent calculation with some other lower-scaling methods such as orbital-optimized M{\o}ller-Plesset perturbation theory.\cite{Lee2018Oct} We do not think that the conclusion of this work will be affected by the choice of single particle basis (i.e., orbitals).

In this section, we will provide the raw data of numerical results that were used in the main text. We will use atomic units for the total energies reported in this section.

\subsection{\ce{H4}, 8-qubit experiment}
We studied a square geometry of \ce{H4} given as
\begin{align*}
\text{H1}&: (0,0,0)\\
\text{H2}&: (0,0,1.23)\\
\text{H3}&: (1.23,0,0)\\
\text{H4}&: (1.23,0,1.23).
\end{align*}
To compute the atomization energy, one needs an energy of a single hydrogen atom. Since Hartree-Fock is an exact approach for a single electron system (e.g., a hydrogen atom), all correlated methods considered in this work should be exact for this. 
For a minimal basis (STO-3G), we used -0.46658185 and for a correlation-consistent quadruple zeta basis (cc-pVQZ) we used -0.499945569 for the hydrogen atom energy.

The classical AFQMC calculations were all performed with a spin-unrestricted Hartree-Fock (UHF) trial wavefunction and we also found that the spin-projection technique (which is often employed to improve the AFQMC results)\cite{purwanto2008eliminating} did not provide any improvement to the AFQMC results. We got -1.96655(4) for STO-3G and -2.10910(8) for cc-pVQZ. CCSD(T) (classical ``gold standard'') was also performed with a UHF reference wavefunction with
energies, -1.961308 (STO-3G) and -2.114275 (cc-pVQZ).

We performed both unpartitioned and partitioned shadow tomography four times for STO-3G and twice for cc-pVQZ.
To get some sense for the convergence of the shadow tomography experiments as a function of the number of sampled Cliffords, we compute the variational energy of the trial wavefunction via
\begin{equation}
E_\text{var} = \frac{\langle \Psi_T|\hat{H}|\Psi_T\rangle}{\langle \Psi_T|\Psi_T\rangle},
\end{equation}
as a function of the number of Cliffords.

\begin{table}[h]
\begin{tabular}{c|c|c|c|c}
$N_\text{Cliffords}$ & repeat 1 & repeat 2 & repeat 3 & repeat 4 \\ \hline
10 &  -1.800644 &  -1.764747 &  -1.813274 &  -1.658202 \\
16 &  -1.823041 &  -1.802192 &  -1.840494 &  -1.730591 \\
28 &  -1.906644 &  -1.839835 &  -1.843326 &  -1.746749 \\
47 &  -1.925654 &  -1.888527 &  -1.860863 &  -1.809656 \\
80 &  -1.909567 &  -1.869456 &  -1.887139 &  -1.846339 \\
136 &  -1.930880 &  -1.902309 &  -1.889992 &  -1.879164 \\
229 &  -1.944249 &  -1.921523 &  -1.903710 &  -1.890947 \\
387 &  -1.947362 &  -1.934682 &  -1.910477 &  -1.901883 \\
652 &  -1.952416 &  -1.939853 &  -1.912790 &  -1.905250 \\
1100 &  -1.955544 &  -1.944651 &  -1.915073 &  -1.909122 \\
1856 &  -1.955028 &  -1.945966 &  -1.909558 &  -1.908038 \\
3129 &  -1.953877 &  -1.947763 &  -1.913386 &  -1.908835 \\
5276 &  -1.954697 &  -1.947323 &  -1.912284 &  -1.909315 \\
8896 &  -1.954930 &  -1.947458 &  -1.913889 &  -1.913068 \\
15000 &  -1.954356 &  -1.948894 &  -1.913894 &  -1.913082 \\
\end{tabular}
\caption {Variational energy of $|\Psi_T\rangle$ from four independent repeated partitioned shadow tomography experiments with a different set of random Cliffords for \ce{H4}, STO-3G (minimal basis). If the experiment was perfect (i.e., no circuit noise), then the variational energy should approach -1.969512.
\label{tab:h4sto3gvar}
}
\end{table}

\begin{table}[h]
\begin{tabular}{c|c|c|c|c}
$N_\text{Cliffords}$ & repeat 1 & repeat 2 & repeat 3 & repeat 4 \\ \hline
10 &  -1.643633 &  -1.798261 &  -1.671065 &  -1.462214 \\ 
16 &  -1.720721 &  -1.848279 &  -1.747911 &  -1.645383 \\ 
28 &  -1.816519 &  -1.911599 &  -1.786704 &  -1.737425 \\ 
47 &  -1.867034 &  -1.920776 &  -1.777655 &  -1.819957 \\ 
80 &  -1.887030 &  -1.901445 &  -1.825170 &  -1.844560 \\ 
136 &  -1.924619 &  -1.930137 &  -1.845217 &  -1.858595 \\ 
229 &  -1.929421 &  -1.933710 &  -1.847781 &  -1.871717 \\ 
387 &  -1.940266 &  -1.936080 &  -1.851352 &  -1.880681 \\ 
652 &  -1.936394 &  -1.937956 &  -1.860513 &  -1.878550 \\ 
1100 &  -1.935905 &  -1.936406 &  -1.875337 &  -1.881012 \\ 
1856 &  -1.938452 &  -1.938114 &  -1.877807 &  -1.884442 \\ 
3129 &  -1.939407 &  -1.939186 &  -1.880363 &  -1.887409 \\ 
5276 &  -1.936669 &  -1.939222 &  -1.882466 &  -1.890464 \\ 
8896 &  -1.937593 &  -1.938921 &  -1.872013 &  -1.888485 \\ 
15000 &  -1.938364 &  -1.939795 &  -1.871097 &  -1.887922
\end{tabular}
\caption {Same as \cref{tab:h4sto3gvar} but for the unpartitioned shadow tomography experiments.
\label{tab:h4sto3gvar2}
}
\end{table}
\begin{table}[h]
\begin{tabular}{c|c|c}
$N_\text{Cliffords}$ & repeat 1 & repeat 2 \\ \hline
10 &  -1.996118 &  -1.658351\\
16 &  -1.988746 &  -1.557607\\
28 &  -2.009853 &  -1.873220\\
47 &  -2.019875 &  -1.976545\\
80 &  -2.026756 &  -1.983726\\
136 &  -2.034241 &  -2.005448\\
229 &  -2.030444 &  -2.045285\\
387 &  -2.051324 &  -2.052698\\
652 &  -2.053210 &  -2.056238\\
1100 &  -2.059021 &  -2.054032\\
1856 &  -2.059920 &  -2.053114\\
3129 &  -2.057736 &  -2.053142\\
5276 &  -2.060762 &  -2.054276\\
8896 &  -2.060786 &  -2.053847\\
15000 &  -2.059437 &  -2.054775\\
\end{tabular}
\caption {Variational energy of $|\Psi_T\rangle$ from four independent repeated partitioned shadow tomography experiments with a different set of random Cliffords for \ce{H4}, cc-pVQZ (a quadruple-zeta basis). If the experiment was perfect (i.e., no circuit noise), then the variational energy should approach -2.069364.
\label{tab:h4ccpvqzvar}
}
\end{table}

\begin{table}[h]
\begin{tabular}{c|c|c}
$N_\text{Cliffords}$ & repeat 1 & repeat 2\\ \hline
10 &  -1.794532 &  -1.961018\\
16 &  -1.864535 &  -1.963510\\
28 &  -1.971853 &  -2.015256\\
47 &  -2.028933 &  -2.025942\\
80 &  -2.022666 &  -2.029521\\
136 &  -2.044745 &  -2.032204\\
229 &  -2.050697 &  -2.036077\\
387 &  -2.055859 &  -2.038768\\
652 &  -2.054068 &  -2.042764\\
1100 &  -2.055576 &  -2.047633\\
1856 &  -2.054740 &  -2.049588\\
3129 &  -2.055636 &  -2.051308\\
5276 &  -2.056442 &  -2.052641\\
8896 &  -2.056741 &  -2.052579\\
15000 &  -2.056641 &  -2.051843\\
\end{tabular}
\caption {Same as \cref{tab:h4ccpvqzvar} but for the unpartitioned shadow tomography experiments.
\label{tab:h4ccpvqzvar2}
}
\end{table}

\begin{table}[h]
\begin{tabular}{c|c|c|c|c}
$N_\text{Cliffords}$ & repeat 1 & repeat 2 & repeat 3 & repeat 4 \\ \hline
10 & -1.96943(5) & -1.98295(6) & -1.96873(6) & -1.9724(1)\\
16 & -1.97376(5) & -1.97385(6) & -1.97175(4) & -1.9672(1)\\
28 & -1.97019(3) & -1.97083(4) & -1.97267(4) & -1.97343(8)\\
47 & -1.97033(2) & -1.96931(3) & -1.97261(4) & -1.97400(7)\\
80 & -1.97016(3) & -1.97398(4) & -1.97061(4) & -1.97038(6)\\
136 & -1.97042(2) & -1.97240(4) & -1.97054(4) & -1.96821(5)\\
229 & -1.97046(2) & -1.97090(2) & -1.96931(4) & -1.96844(5)\\
387 & -1.97019(2) & -1.97076(2) & -1.97010(4) & -1.96831(5)\\
652 & -1.97030(2) & -1.97013(2) & -1.96929(4) & -1.96861(4)\\
1100 & -1.96928(2) & -1.96958(2) & -1.96931(4) & -1.96882(5)\\
1856 & -1.96942(2) & -1.96964(1) & -1.96974(4) & -1.96909(5)\\
3129 & -1.96914(2) & -1.96948(2) & -1.96933(4) & -1.96922(4)\\
5276 & -1.96879(2) & -1.96947(2) & -1.96914(4) & -1.96944(5)\\
8896 & -1.96877(2) & -1.96959(2) & -1.96918(4) & -1.96952(4)\\
15000 & -1.96877(2) & -1.96964(2) & -1.96922(4) & -1.96941(4)
\end{tabular}
\caption {AFQMC energy using $|\Psi_T\rangle$ from four independent repeated partitioned shadow tomography experiments with a different set of random Cliffords for \ce{H4}, STO-3G (minimal basis). The exact ground state energy is -1.969512. The numbers in parentheses indicate the statistical error of the AFQMC energy.
\label{tab:h4sto3gafqmc}
}
\end{table}

\begin{table}[h]
\begin{tabular}{c|c|c|c|c}
$N_\text{Cliffords}$ & repeat 1 & repeat 2 & repeat 3 & repeat 4 \\ \hline
10 & -2.0058(1) & -1.97058(9) & -1.9712(1) & -1.9823(2)\\
16 & -1.9907(1) & -1.96982(8) & -1.97094(9) & -1.9869(1)\\
28 & -1.98318(7) & -1.96711(4) & -1.97036(9) & -1.97288(6)\\
47 & -1.97642(5) & -1.96859(3) & -1.9823(1) & -1.97291(6)\\
80 & -1.97430(4) & -1.97010(5) & -1.9833(1) & -1.96990(5)\\
136 & -1.97131(3) & -1.96846(3) & -1.97343(8) & -1.97025(6)\\
229 & -1.97114(2) & -1.96934(3) & -1.97253(8) & -1.96970(6)\\
387 & -1.96995(2) & -1.97006(3) & -1.97059(8) & -1.96981(6)\\
652 & -1.96982(3) & -1.96995(3) & -1.97024(7) & -1.96980(7)\\
1100 & -1.96975(3) & -1.97054(3) & -1.96955(7) & -1.96958(7)\\
1856 & -1.96940(3) & -1.97017(3) & -1.96886(7) & -1.96975(7)\\
3129 & -1.96926(3) & -1.97013(3) & -1.96884(7) & -1.96984(7)\\
5276 & -1.96940(3) & -1.96999(3) & -1.96931(7) & -1.96968(7)\\
8896 & -1.96950(3) & -1.97011(3) & -1.96918(8) & -1.96954(7)\\
15000 & -1.96952(3) & -1.97022(3) & -1.96943(7) & -1.96930(7)
\end{tabular}
\caption {Same as \cref{tab:h4sto3gafqmc} but for the unpartitioned shadow tomography experiments.
\label{tab:h4sto3gafqmc2}
}
\end{table}
\begin{table}[h]
\begin{tabular}{c|c|c}
$N_\text{Cliffords}$ & repeat 1 & repeat 2  \\ \hline
10 & -2.10573(9) & -2.1461(3)\\
16 & -2.10766(9) & -2.1214(5)\\
28 & -2.1095(1) & -2.1344(3)\\
47 & -2.1107(2) & -2.1214(1)\\
80 & -2.11063(5) & -2.1313(2)\\
136 & -2.11039(6) & -2.1220(1)\\
229 & -2.11044(6) & -2.11312(5)\\
387 & -2.11120(7) & -2.11141(4)\\
652 & -2.11026(7) & -2.11176(7)\\
1100 & -2.11090(4) & -2.11105(4)\\
1856 & -2.11067(3) & -2.11131(4)\\
3129 & -2.11055(6) & -2.11120(5)\\
5276 & -2.11105(4) & -2.11090(4)\\
8896 & -2.11119(5) & -2.11092(6)\\
15000 & -2.11081(3) & -2.11098(4)
\end{tabular}
\caption {AFQMC energy using $|\Psi_T\rangle$ from four independent repeated partitioned shadow tomography experiments with a different set of random Cliffords for \ce{H4}, cc-pVQZ (a quadruple-zeta basis). The exact ground state energy is -2.11216599. The numbers in parentheses indicate the statistical error of the AFQMC energy.
\label{tab:h4ccpvqzafqmc}
}
\end{table}

\begin{table}[h]
\begin{tabular}{c|c|c}
$N_\text{Cliffords}$ & repeat 1 & repeat 2 \\ \hline
10 & -2.1188(2) & -2.1070(1)\\
16 & -2.1146(1) & -2.1080(1)\\
28 & -2.10942(9) & -2.11169(9)\\
47 & -2.10951(6) & -2.11108(7)\\
80 & -2.1111(1) & -2.11219(7)\\
136 & -2.11100(4) & -2.11064(6)\\
229 & -2.11105(4) & -2.11218(6)\\
387 & -2.11069(3) & -2.11197(7)\\
652 & -2.11068(4) & -2.11159(8)\\
1100 & -2.11048(4) & -2.11180(5)\\
1856 & -2.1109(1) & -2.11206(6)\\
3129 & -2.11092(6) & -2.11198(5)\\
5276 & -2.11015(3) & -2.11186(5)\\
8896 & -2.11045(3) & -2.11220(5)\\
15000 & -2.11040(4) & -2.11182(5)
\end{tabular}
\caption {Same as \cref{tab:h4ccpvqzafqmc} but for the unpartitioned shadow tomography experiments.
\label{tab:h4ccpvqzafqmc2}
}
\end{table}

The corresponding variational energies are shown in \cref{tab:h4sto3gvar} and \cref{tab:h4sto3gvar2} for a minimal basis set (STO-3G) varying the number of Clifford circuits. Using these trial wavefunctions we computed the phaseless AFQMC energies (i.e., QC-AFQMC energies) as shown in \cref{tab:h4sto3gafqmc} and \cref{tab:h4sto3gafqmc2}.
There is significant variation in the variational energy depending on the number of Cliffords and whether one uses partitioned shadow tomography or not. Nonetheless, the subsequent AFQMC energy is nearly converged with respect to the number of Cliffords at 15000 and run-to-run variation is negligible. We observe essentially the same qualitative results in the case of cc-pVQZ as shown in \cref{tab:h4ccpvqzafqmc} and \cref{tab:h4ccpvqzafqmc2}.

\subsection{\ce{N2}, 12-qubit experiment}
For \ce{N2}, we performed only one set of partitioned shadow tomography experiments with a total of 15000 Cliffords because we observed that our final AFQMC energy varies very slightly run-to-run in the case of \ce{H4}. We used a correlation-consistent triple-zeta basis, cc-pVTZ.\cite{Dunning1989}
The classical AFQMC calculations done with UHF trial wavefunctions and the spin-projection technique did not change the results discussed here. Similarly, we used UHF reference states for CCSD(T) calculations.
Here, we provide the raw data which was used in \cref{fig:fig3} (a).
Our exact results are obtained from HCI where the second-order perturbation correction was found to be smaller than 0.002 a.u. We believe that these ``exact'' results are converged with enough precision that these numbers can be used as a benchmark for this system.

\begin{table}[h]
\begin{tabular}{c|c|c|c|c|c}
R(\AA) & Exact & CCSD(T) & Quantum trial & AFQMC & QC-AFQMC \\ \hline 
1.000& -109.366398& -109.365383& -109.017231 & -109.3672(3) & -109.36697(7)\\
1.125& -109.399981& -109.398412& -109.043176 & -109.4003(3) & -109.40094(7)\\
1.250& -109.360887& -109.355280& -109.000672 & -109.3603(4) & -109.36085(8)\\
1.500& -109.233325& -109.215012& -108.874636 & -109.2342(3) & -109.23109(9)\\
1.750& -109.132826& -109.110942& -108.808418 & -109.1408(2) & -109.13325(8)\\
2.000& -109.080654& -109.066772& -108.790143 & -109.0939(2) & -109.08341(7)\\
2.250& -109.061147& -109.053758& -108.788486 & -109.07392(8) & -109.06177(7)
\end{tabular}
\caption {Raw data for \ce{N2} potential energy surface for seven bond distances ($R$).
Note that the energy of our quantum trial here is obtained from a single set of experiment which may vary significantly run-to-run.
\label{tab:n2ccpvtz}
}
\end{table}

\subsection{Diamond, 16-qubit experiment}
For diamond, we used the GTH-PADE pseudopotential\cite{Goedecker1996Jul} and the DZVP-GTH basis.\cite{VandeVondele2007Sep} Only the $\Gamma$-point was considered in the Brillouin zone sampling and the computational unit cell consists of only two carbon atoms.
We used spin-restricted HF (RHF) trial wavefunctions for classical AFQMC calculations and CCSD(T) also employed RHF reference states. The ``exact'' results are obtained from HCI and the second-order perturbation correction was found to be smaller than 0.0001 a.u. These results should be good as reference data.
We took a total of 50000 Clifford samples to perform a set of partitioned shadow tomography experiments at all lattice constants considered. 
In \cref{tab:diamonddzvp}, we present the raw data used for \cref{fig:fig3} (b).

\begin{table}[h]
\begin{tabular}{c|c|c|c|c|c}
R(\AA) & Exact & CCSD(T) & Quantum trial & AFQMC & QC-AFQMC \\ \hline 
2.880&  -9.545911&  -9.546464&  -9.121081 & -9.5415(1) & -9.54582(5)\\
3.240& -10.229155& -10.230100&  -8.625292 & -10.2241(3) & -10.23051(7)\\
3.600& -10.560477& -10.562229& -10.277938 & -10.5525(2) & -10.55861(8)\\
3.960& -10.700421& -10.703884& -10.368882 & -10.6869(2) & -10.6949(1)\\
4.320& -10.744089& -10.751103& -10.222206 & -10.7177(3) & -10.73701(9)
\end{tabular}
\caption {Raw data for the diamond cold curve for five lattice constants ($R$).
Note that the energy of our quantum trial here is obtained from a single set of experiment which may vary significantly run-to-run. Note that these energies include the Madelung constant.
\label{tab:diamonddzvp}
}
\end{table}

\subsection{Quantum Circuit Details}

\begin{table}[]
  \begin{tabular}{@{}llllll@{}}
    \cmidrule(r){1-5}
    Experiment               & \# Qubits & \# CZ Gates (State Prep) & \# CZ Gates (Total) & Circuit Depth \\ \cmidrule(r){1-5}\hline
    Hydrogen (Partitioned)                 & 8         & 36                                               & 66                  & 52                    \\
    Hydrogen (Unpartitioned) & 8         & 36                                                & 99                  & 67                    \\
    Nitrogen                 & 12        & 22                                                & 92                  & 53                    \\
    Diamond                  & 16        & 34                                                & 160                 & 65                    \\ 
    \cmidrule(r){1-5}
  \end{tabular}
  \caption{Resource counts for the QC-AFQMC experiments realized in this work.
  \label{tab:resouce1}
  }
\end{table}

\begin{table}[]
  \begin{tabular}{@{}llll@{}}
    \cmidrule(r){1-3}
    Experiment & Reference               & \# Qubits & \# 2q Gates              \\
    \hline
    BeH$_{2}$  & \cite{kandala2017hardware}  & 6          & 5 ($U_{\mathrm{ENT}}$)               \\
    H$_{2}$O   & \cite{nam2020ground}        & 5          & 6 ($XX(\theta)$) \\
    Hydrogen   & \cite{google2020hartree}    & 12         & 72 ($\sqrt{i\textsc{swap}}$)               \\
    Diazene    & \cite{google2020hartree}    & 10         & 50 ($\sqrt{i\textsc{swap}}$)              \\
    Hubbard, interacting (8-site) & \cite{arute2020observation} & 16        & 608 ($\sqrt{i\textsc{swap}}$) \\
    Hubbard, non-interacting (8-site) & \cite{arute2020observation}     & 16        & 1568  ($\sqrt{i\textsc{swap}}$) \\
    \cmidrule(r){1-3}
  \end{tabular}
  \caption{Resource estimates from prior fermionic simulations using gate model quantum computers on more than four qubits. For the two Hubbard model experiments we distinguish between dynamics simulated for an interacting versus a non-interacting model.  $N=8$ indicates an eight site linear lattice with open boundary conditions. $U_{\mathrm{ENT}}$ is a nearest-neighbor cross-resonance style gate and $XX(\theta)$ is a $\mathrm{exp}(-i\theta \sigma^{i}_{x} \sigma^{j}_{x}/2)$. As far as we are aware, these are the largest simulations using a gate-model quantum computer targeting fermionic ground states or dynamics.
  \label{tab:resouce2}
  }
\end{table}
In this section we describe the construction of the particular circuits we used in our experiments. 
In \cref{tab:resouce1} and \cref{tab:resouce2},
we summarize the quantum resource usage in our experiments and other prior works.

The circuits to be applied have two parts: the part that prepares the superposition of the trial wavefunction and the zero state, and the shadow tomography part that implements the measurement operator.

Our trial wave functions are perfect pairing states, followed by some number preserving fermionic gates in the case of the eight qubit experiment.
Because the state we want to prepare is
\begin{equation}
\Ket{\tau} = \left(\Ket{0} + \Ket{\Psi_T}\right) / \sqrt{2},
\end{equation}
it is sufficient to prepare 
\begin{equation}
\left(\Ket{0} + \Ket{\mathrm{PP}(\boldsymbol \theta)}\right) / \sqrt{2},
\end{equation}
where 
\begin{equation}
\Ket{\mathrm{PP}(\boldsymbol \theta)}
=
\bigotimes_{i=1}^{N/4} \Ket{\mathrm{PP}(\theta_i)}
\end{equation}
and $N$ is the number of spin orbitals.
We do this by creating a state
\begin{equation}
\left(\Ket{0} + {\Ket{1000}}^{\otimes N/4}\right) / \sqrt{2}
\end{equation}
using a single-qubit Hadamard and a ladder of CNOT and SWAP gates.
Then for each set of 4 qubits corresponding to a pair of spatial orbitals we prepare 
\begin{align}
\Ket{\mathrm{PP}(\theta)}
=
\cos(\theta) \Ket{1100}
+
\sin(\theta) \Ket{0011}
\propto
\CNOT_{1, 2}
\CNOT_{3, 4}
{\left(i\mathrm{SWAP}_{1, 3}\right)}^{\theta}
\Ket{1000}
,
\end{align}
where the CNOTs and iSWAP gates leave the zero part of the state unchanged. See the portion of  \fig{fig2} (a) labelled "perfect pairing" for a circuit diagram illustrating this step. \fig{fig2} (a) also shows a circuit diagram of the aforementioned additional number preserving gates used in the eight qubit experiment. The perfect pairing states, as well as the number preserving gates, are discussed from a quantum chemical perspective in \app{trial}. 


Now we discuss how to implement the measurement operators.
As discussed in Sec.~\ref{app:global-stabilizer-measurements}, the measurement operators
have the form
\begin{equation}
  G(I, \Gamma, \Delta)
  =
  \prod_{i \in I} H_i P_i^{\Gamma_{i, i}}
  \prod_{\substack{i \in I \\ j \in I: j \neq i}} \CZ_{i, j}^{\Gamma_{i, j}}
  \prod_{\substack{
      i \in I\\
      j \notin I: j > i
    }
  } \CX_{i, j}^{\Delta_{i, j}}
  .
\end{equation}
Let $\tilde{\Gamma} = \Gamma + \Delta$.
We can rewrite $G$ as
\begin{equation}
  G(I, \Gamma, \Delta)
  =
  H^{\otimes n}
  \prod_{i \in I} P_i^{\Gamma_{i, i}}
  \prod_{i, j} \CZ_{i, j}^{\tilde{\Gamma}_{i, j}}
  \prod_{i \notin I} H_i
  ,
\end{equation}
i.e.,
a CZ layer sandwiched by two layers of single-qubit gates.
Maslov and Roetteler~\cite{maslov2018shorter} showed that a CZ layer followed by
complete reversal of the qubits can be implemented using a circuit of $2n + 2$
CNOT layers (plus intervening layers of single qubit powers of P).
Because the CZ layer in the circuit for $G$ is followed only by single-qubit
gates and measurement in the computational basis, the reversal of qubits can be
easily undone in post-processing.
Thus the shadow tomography circuits have a 2-qubit gate depth of at most $2n +
2$.
This is a significant improvement over using the full Clifford group for shadow tomography; 
the best known circuit for a general Clifford has 2-qubit depth $9n$.
\cite{bravyi2020hadamardfree}
Furthermore, the CZ circuits have the additional properties that they contain
only four unique CNOT layers and that they act only along a line, which are
advantageous for calibration and qubit mapping, respectively.

\section{Outlook on Potential Quantum Advantage}\label{app:quantumadvantage}
In the typical electronic structure context, quantum advantage is focused on the approximation of the ground state energy.
In this outlook, we consider the potential for quantum advantage in this general sense, as well as for the specific quantum subroutine used in our QC-AFQMC algorithm, namely the overlap evaluation.
We explain our understanding here of the current computational scaling and limits of our proposed approach for the overlap evaluation and the path towards the first ``practical'' quantum advantage.

{\it System size scaling}.
In general, we expect the overlap between $\langle \Psi_T | \phi\rangle$ to
approach zero exponentially quickly as the system size increases.
For example, the typical overlap value of the walker wavefunction with a simple trial wavefunction can be as small as $10^{-5}$ for 16 atoms, $10^{-16}$ for 54 atoms, and $10^{-38}$ for 128 atoms under periodic boundary conditions.\cite{malone_isdf} These examples suggest that the system size scaling consideration is not just an asymptotic consideration but is practically relevant for system sizes that one may wish to study in the near future.
Performing AFQMC requires evaluating these overlaps to a fixed relative precision. Therefore, as the system size increases towards the thermodynamic limit, we would expect that QC-AFQMC formally requires exponentially more measurements to maintain the relative precision.

In order to address the challenges due to this scaling, QC-AFQMC might need to be developed beyond the formulation used in our experiment. For example, using more sophisticated wavefunction forms for $|\phi\rangle$ than a single Slater determinant could allows one to maintain good overlap between $|\Psi_T\rangle$ and $|\phi\rangle$. Again, as long as $|\phi\rangle$ can be prepared efficiently on a quantum computer, one can efficiently estimate the overlap $\langle \Psi_T | \phi\rangle$ to fixed additive error using the Hadamard test. In some cases, these overlaps might still be too small as a consequence of the QMA-Hardness of the electronic structure problem.\cite{Schuch2009} However, the onset of this sort of exponential scaling would also render intractable other quantum computing algorithms such as energy estimation via quantum phase estimation.\cite{Abrams1999} The reason for this is because such approaches have a cost that is inversely proportional to the overlap between the target eigenstate of interest and the initial state. Thus, if one can make quantum phase estimation efficient by preparing a suitable initial state, we are optimistic that one can use parameterized versions of those states as the initial $|\phi\rangle$ in QC-QMC in order to avoid the problem of vanishing overlaps. We note that VQE is also expected to face similar difficulties in the worst case since no polynomial scaling circuit ansatz is able to prepare ground states of the most challenging instances of the electronic structure problem to target precision.

Alternatively, one could pursue strategies for controlling the sign problem which do not require computing the global wavefunction overlaps to a high precision directly. Classically, the exponential decay of these overlap values with respect to system size for single Slater determinant walkers is numerically well handled by computing the $\log$ of the overlap value directly and working only with the overlap ratio when performing the AFQMC calculations. While that particular strategy seems difficult to implement on a quantum computer, it seems reasonable that one could leverage the finite correlation length of physical systems to avoid the need for an exponentially growing number of measurements. More specifically, our virtual correlation technique allows for choosing a relatively small physical space to treat with the quantum processor while computing the correlation energy in a much larger space. The use of such a small physical space (known as an active space in quantum chemistry) can be rigorously justified for systems with a finite correlation length. The typical wavefunction overlaps under this approach would therefore be (at worst) exponentially small in a quantity related to the correlation length rather than the size of the system. Furthermore, it is often possible to keep the physical space small by identifying a reduced set of physically relevant degrees of freedom. In practice, the combination of these facts will help us maintain overlaps much larger than we would expect in the most general cases.

{\it Quantum advantage in the overlap estimation}.
A related but independently interesting question is whether there is a potential for quantum advantage with regards to the specific task of estimating the overlap up to an additive error between some quantum state and an arbitrary walker wavefunction (a single Slater determinant in our particular experiments).
Although the use of shadow tomography is guaranteed to be efficient for this task in terms of the number of measurements, the classical post-processing used in our shadow tomography experiments was performed with an exponential overhead incurred by enumerating all possible determinants in the Hilbert space (see 
\app{shadow_tomo_overlap} and \cref{app:vmc}). One open question raised by our work is whether there is a way to remove this exponential overhead in the classical post-processing of shadow tomography for QC-AFQMC, possibly by using a different ensemble of random unitaries. Building on \citen{Zhao2020-fv}'s fermionic shadow tomography seems promising in this regard. Even if the answer is no, one does not need to use shadow tomography; using the Hadamard test, one can obtain the overlaps up to additive error efficiently without any problematic classical post-processing. Thus, in general, one can estimate these overlaps up to an additive error in a fashion that is entirely efficient. One could also pursue a version of QC-QMC that avoids this obstacle by using walkers that are particularly well suited for use with shadow tomography, e.g., composed of a linear combination of stabilizer states (states generated by Clifford circuits). The Green's function Monte Carlo method is one example of this (as the walker wavefunctions are computational basis states).

We employed the perfect pairing (PP) wavefunction as a workhorse in all our experiments. While to the best of our knowledge there is no efficient classical algorithm that can compute the overlap between a PP state and an arbitrary single Slater determinant {\it exactly}, there is an efficient classical algorithm (see \cref{app:vmc}) that can approximate this quantity up to some additive error. Therefore, we can assert that there is no quantum advantage in using PP trial wavefunctions in QC-AFQMC. On the other hand, more complex states such as the one used in our \ce{H4} experiment (i.e., PP state with hardware efficient layers), other hardware-efficient wavefunctions, some variants of the unitary coupled-cluster (UCC) wavefunction (see \cref{app:trial}), and the two-dimensional multiscale entanglement renormalization (2D-MERA) wavefunction may be good candidates for seeking a quantum advantage in the estimation of overlaps. This is due to the fact that no known classical algorithms (including the one described in \cref{app:vmc}) efficiently yield the overlap of these wavefunctions (up to an additive error) with an arbitrary Slater determinant, or indeed, a computational basis state. Overlaps between all these states and a single Slater determinant can be approximated efficiently up to additive error on the quantum computer using the Hadamard test. Overlaps of these states with stabilizer states (including computational basis states) can be approximated efficiently using existing shadow tomography techniques.

{\it Quantum advantage in the ground state energy computation}.
When the number of electrons that we consider is not too large, it is possible to assume that the measurement overhead due to the vanishing overlap may not be a practical concern. With our virtual correlation technique, we can maintain a good overlap value within the active space while producing accurate energies overall. Given this, we are optimistic about routes to achieve quantum advantage in fermionic ground state simulation through the QC-AFQMC algorithm.
The aforementioned complex quantum states such as hardware efficient ansatze, UCC and 2D-MERA can be good candidates for trial wavefunctions although the relevance of 2D-MERA for chemistry simulations is yet to be seen.
An important consideration here is how one actually obtains wavefunction parameters of those complex quantum states. One may optimize them using the variational quantum algorithm or one may take states that can be efficiently optimized classically. For the latter case, it seems likely that approximating the overlap between these states and an arbitrary Slater determinant up to additive error is difficult despite the fact that some of them can be optimized efficiently using classical algorithms.

We hope to observe quantum advantage either in the overlap estimation or in the ground state energy computation using QC-AFQMC and believe that continued advancement along this direction will lead us to one of the first realizations of practical quantum advantage in NISQ fermionic simulations.




\end{document}